\documentclass[prd,onecolumn,nopacs,nofootinbib]{revtex4}

\usepackage{amsmath}
\usepackage{graphicx}
\usepackage{amsfonts}
\usepackage{latexsym}
\usepackage{bbold}
\usepackage{wasysym}
\usepackage{calligra}
\usepackage{float}
\usepackage{ulem}
\usepackage{inputenc}
\usepackage{xspace}
\usepackage{url}
\usepackage{epstopdf}
\usepackage{tikz}

\newcommand{\be}{\begin{equation}}
\newcommand{\ee}{\end{equation}}
\newcommand{\beq} {\begin{equation}}
\newcommand{\eeq} {\end{equation}}
\newcommand{\ba}{\begin{eqnarray}}
\newcommand{\ea}{\end{eqnarray}}

\newtheorem{mydef}{Definition}
\newtheorem{corollary}{Corollary}

\newtheorem{theo}{Theorem}

\begin{document}

	\title{The Full Quadratic Metric-Affine Gravity (Including Parity Odd Terms): Exact solutions for the Affine-Connection}
	
	\author{Damianos Iosifidis}
	\affiliation{Institute of Theoretical Physics, Department of Physics
		Aristotle University of Thessaloniki, 54124 Thessaloniki, Greece}
	\email{diosifid@auth.gr}
	
	\date{\today}
	\begin{abstract}
		
We consider the most general Quadratic Metric-Affine Gravity setup in the presence of generic matter sources with non-vanishing hypermomentum. The gravitational action consists of all $17$ quadratic invariants (both parity even and odd) in torsion and non-metricity as well as their mixings, along  with the terms that are linear in the curvature namely the Ricci scalar and the totally antisymmetric Riemann piece. Adding also a matter sector to the latter we first obtain the field equations for the generalized quadratic Theory. Then, using a recent Theorem,  we successfully find the exact form of the affine connection under some quite general non-degeneracy conditions. Having obtained the exact and unique solution of the affine connection we subsequently derive the closed forms of spacetime torsion and non-metricity and also recast the metric field equations into  a GR form with modified source terms that are quadratic in the hypermomentum and linear in its derivatives. We also study the vacuum quadratic Theory and prove that in this instance, or more generally for vanishing hypermomentum, the connection becomes the Levi-Civita one. Therefore, we also find exactly to  what does the quadratic vacuum Theory correspond to. Finally, we generalize our result even further  and also discuss the physical consequences and applications of our study.
		
	\end{abstract}
	
	\maketitle
	
	\allowdisplaybreaks
	
	
	\tableofcontents
	
	\section{Introduction}
	\label{intro}

	General Relativity (GR) is one of the pillars of modern Theoretical Physics. Apart from its rigorous mathematical formulation and the solid predictions it provides, the geometrization of gravity that is associated to it makes the latter particularly appealing. However, GR has been challenged by many observational data and the very fact that it cannot be quantized  makes it clear that the latter is not the final answer to the puzzle of gravitation. Driven mostly by Cosmological data many alternative Theories of gravity have been proposed over the years \cite{clifton2012modified,CANTATA:2021ktz}. From the great plethora of alternatives to GR, of particular interest are those extensions/modifications that are in this spirit of geometrization of gravitational phenomena that GR preaches. In this direction, one can extend the geometric arena upon which GR is based by enriching it with non-Riemannian structures as well. Namely, one can alleviate the Riemannian constraints of vanishing torsion and non-metricity and be in the realm of non-Riemannian geometry \cite{eisenhart2012non}. In this extended geometry, the linear connection is, in general, neither symmetric (i.e. it processes torsion) nor is it compatible with the metric (non-metricity). The gravity Theory that is formulated on such a non-Riemannian manifold is widely known as Metric-Affine Gravity (MAG) \cite{hehl1995metric,iosifidis2019metric}.
	
	In MAG one starts off with a general affine connection that is independent of the metric and it possesses both torsion and non-metricity. The field equations are then obtained by independent variations with respect to the metric and the affine connection. There are many reasons that make MAG an appealing gravity Theory\footnote{See \cite{hehl1995metric} for an extended list of motives for studying MAG.} but probably the most astonishing one is the link it provides in relation to the microstructure of matter \cite{hehl1991spacetime,puetzfeld2008probing}.
	In recent years, there is an ever-increasing interest in the MAG formulation
	\cite{iosifidis2018raychaudhuri,iosifidis2019exactly,iosifidis2019scale,vitagliano2011dynamics,sotiriou2007metric,capozziello2010metric,percacci2020new,Jimenez:2020dpn,BeltranJimenez:2019acz,delhom2018observable,aoki2019scalar,Cabral:2020fax,Ariwahjoedi:2020wmo,Ariwahjoedi:2021yth,Yang:2021fjy,hohmann2020metric,bahamonde2020new} with a special emphasis given in Cosmological applications 
	 \cite{Iosifidis:2021fnq,iosifidis2020cosmic,Iosifidis:2020gth,Iosifidis:2021iuw,jimenez2016spacetimes,beltran2017modified,kranas2019friedmann,barragan2009bouncing,shimada2019metric,Yang:2021fjy,kubota2021cosmological,Mikura:2020qhc,Mikura:2021ldx,Babourova:2002fn,Puetzfeld:2001hk,Minkevich:1998cv}\footnote{For the importance of non-Riemannian effects in the early Universe see \cite{Puetzfeld:2004yg} and references therein. }. Let us also mention that	
	MAG  has some promising features in regards to the Quantization of Gravity (see ref \cite{Percacci:2020bzf}). From the above discussion it becomes clear that MAG is worth studying and has a lot to offer in relation to our understanding of Gravity and the effects of microstructure of matter. Our concern here will be to study the most general MAG Theory consisting of all possible (both parity even and parity odd) quadratic invariants\footnote{Here we wish to generalize the findings of \cite{Iosifidis:2021tvx} where only the parity even quadratic  terms were included.} in torsion and non-metricity along with terms that are linear in the curvature. That is our gravitational action will be comprised of the usual Einstein-Hilbert term (i.e Ricci scalar) and the $17$ quadratic invariants of torsion and non-metricity. To this action we shall also add matter which will, in general, depend also on the connection, that is it will possess a non-vanishing hypermomentum. Our task will then be to find the general form of the connection for the quadratic Theory in the presence of matter but for the vacuum case as well.  
	
	More precisely, the paper is organized as follows. Firstly, we set up conventions and the definitions we are going to use throughout and briefly discuss some geometrical as well as physical aspects of MAG. We then consider the general $17$ parameter quadratic MAG Theory in the presence of matter. After varying independently with respect to the metric and the affine connection we derive the field equations of the quadratic Theory. Then, focusing on the connection field equations, by performing a simple trick and using a recent result (see \cite{Iosifidis:2021tvx}) on how to solve complicated linear tensor equations, we obtain the exact a unique solution of the affine connection. In our step by step proof, we show that the latter is expressed in terms of the usual Levi-Civita part plus modifications coming from various combinations/dualizations and contractions of the hypermomentum tensor. Subsequently we compute the associated forms of torsion and non-metricity as they emerge from the microstructure of matter (i.e. hypermomentum). With these results we are then able to perform a post-Riemannian expansion on the metric field equations and recast them as an effective GR with modified source terms emerging from the intrinsic characteristics of matter. Finally, by switching-off the sources we show that the vacuum version of the quadratic Theory, despite its complexity, is always equivalent to GR given that there are no degeneracies. The same result continuous to hold true  also for matter that does not couple to the connection (i.e vanishing hypermomentum). We then generalize our result even further and finally conclude  and discuss future applications.

		\section{The Setup}
	We start off by considering a $4-dim$ differentiable  manifold endowed with a symmetric metric and an independent affine connection ($\mathcal{M}$, g, $\nabla$). We will use the definitions and notations of \cite{Iosifidis:2020gth} and therefore will go through them rather briefly here. On this non-Riemannian manifold endowed with the metric $g_{\mu\nu}$ and an independent affine connection with components $\Gamma^{\lambda}_{\;\;\;\mu\nu}$, we define the curvature, torsion and non-metricity tensors according to
	\beq
	R^{\mu}_{\;\;\;\nu\alpha\beta}:= 2\partial_{[\alpha}\Gamma^{\mu}_{\;\;\;|\nu|\beta]}+2\Gamma^{\mu}_{\;\;\;\rho[\alpha}\Gamma^{\rho}_{\;\;\;|\nu|\beta]} \label{R}
	\eeq
	\beq
	S_{\mu\nu}^{\;\;\;\lambda}:=\Gamma^{\lambda}_{\;\;\;[\mu\nu]}
	\eeq
	\beq
	Q_{\alpha\mu\nu}:=- \nabla_{\alpha}g_{\mu\nu}
	\eeq
	Of great importance is the  the so-called distortion tensor which measures how much
	the general affine connection $\Gamma^{\lambda}_{\;\;\;\mu\nu}$ deviates away from the Levi-Civita one \cite{schouten1954ricci}, namely
	\begin{gather}
	N^{\lambda}_{\;\;\;\;\mu\nu}:=\Gamma^{\lambda}_{\;\;\;\mu\nu}-\widetilde{\Gamma}^{\lambda}_{\;\;\;\mu\nu}=
	\frac{1}{2}g^{\alpha\lambda}(Q_{\mu\nu\alpha}+Q_{\nu\alpha\mu}-Q_{\alpha\mu\nu}) -g^{\alpha\lambda}(S_{\alpha\mu\nu}+S_{\alpha\nu\mu}-S_{\mu\nu\alpha}) \label{N}
	\end{gather}
	where $\widetilde{\Gamma}^{\lambda}_{\;\;\;\mu\nu}$ is the usual Levi-Civita connection calculated only by the metric and its first derivatives. Once the distortion is given, torsion and non-metricity are readily computed through (see for instance \cite{iosifidis2019metric})
	\beq
	S_{\mu\nu\alpha}=N_{\alpha[\mu\nu]}\;\;,\;\;\; Q_{\nu\alpha\mu}=2 N_{(\alpha\mu)\nu} \label{QNSN}
	\eeq
	Out of torsion we can construct a vector as well as a pseudo-vector. Our definitions for the torsion vector and pseudo-vector are
	\beq
	S_{\mu}:=S_{\mu\lambda}^{\;\;\;\;\lambda} \;\;, \;\;\;
	t_{\mu}:=\epsilon_{\mu\alpha\beta\gamma}S^{\alpha\beta\gamma} 
	\eeq
	respectively. Note that the former is defined for any dimension while the latter only for the $n=4$ case we are considering here. Continuing with non-metricity, we define the Weyl  and the second non-metricity vector according to
	\beq
	Q_{\alpha}:=Q_{\alpha\mu\nu}g^{\mu\nu}\;,\;\; q_{\nu}=Q_{\alpha\mu\nu}g^{\alpha\mu}
	\eeq
	Finally,  from the Riemann tensor we can construct the three contractions
	\beq
	R_{\nu\beta}:=R^{\mu}_{\;\;\nu\mu\beta}	
	\eeq
	\beq
	\hat{R}_{\alpha\beta}:=R^{\mu}_{\;\;\mu\alpha\beta}	
	\eeq
	\beq
	\breve{R}^{\mu}_{\;\;\beta}:=R^{\mu}_{\;\;\nu\alpha\beta}	g^{\nu\alpha}
	\eeq
	
	As usual, the first one is the Ricci tensor (which is not symmetric in general), the second one is the homothetic curvature and the last one is the the co-Ricci tensor. Note that the first two aforementioned tensors can be formed without the use of any metric while for the latter a metric is required. As for the generalized Ricci scalar, the latter is still uniquely defined since
	\beq
	R:=g^{\mu\nu}R_{\mu\nu}=-g^{\mu\nu}\breve{R}_{\mu\nu} \;\; ,\;\;\; g^{\mu\nu}\hat{R}_{\mu\nu}=0
	\eeq
	
	Of prominent importance and of great use in our proceeding discussion is the decomposition ($\ref{N}$).
	Indeed,  by virtue of ($\ref{N}$) each quantity can be split into its Riemannian part (i.e. computed with respect to the Levi-Civita connection) plus non-Riemannian contributions. For instance inserting the connection decomposition ($\ref{N}$) into the definition ($\ref{R}$) we obtain for the Riemann tensor \footnote{Quantities with\;  $\widetilde{}$\;  will always denote Riemannian parts  unless otherwise stated.} 
	\beq
	{R^\mu}_{\nu \alpha \beta} = \widetilde{R}^\mu_{\phantom{\mu} \nu \alpha \beta} + 2 \widetilde{\nabla}_{[\alpha} {N^\mu}_{|\nu|\beta]} + 2 {N^\mu}_{\lambda|\alpha} {N^\lambda}_{|\nu|\beta]} \,, \label{decomp}
	\eeq
	The last decomposition is very useful and we are going to be using it later on  in order to express the metric field equations of our Theory in Einstein-like form plus modified matter sources. For instance, with the use of the above the post-Riemannian expansion of the Ricci scalar reads
	\begin{gather}
	R=\tilde{R}+ \frac{1}{4}Q_{\alpha\mu\nu}Q^{\alpha\mu\nu}-\frac{1}{2}Q_{\alpha\mu\nu}Q^{\mu\nu\alpha}    -\frac{1}{4}Q_{\mu}Q^{\mu}+\frac{1}{2}Q_{\mu}q^{\mu}+S_{\mu\nu\alpha}S^{\mu\nu\alpha}-2S_{\mu\nu\alpha}S^{\alpha\mu\nu}-4S_{\mu}S^{\mu} \nonumber \\ +2 Q_{\alpha\mu\nu}S^{\alpha\mu\nu}+2 S_{\mu}(q^{\mu}-Q^{\mu}) +\tilde{\nabla}_{\mu}(q^{\mu}-Q^{\mu}-4S^{\mu})
	\end{gather}

Before closing this section let us introduce some useful definitions that we are going to use throughout.
 \begin{mydef}Consider the components $N_{\alpha\mu\nu}$ of a rank-3 tensor field.\footnote{Of course these definition hold true for arbitrary rank $3$ tensors but in our discussion here this tensor will be the distortion.} We define the $1^{st}$, $2^{nd}$ and $3^{rd}$ contractions of $N_{\alpha\mu\nu}$ according to
	\beq
	N^{(1)}_{\mu}:=N_{\alpha\beta\mu}g^{\alpha\beta}\;\;, \;\; N^{(2)}_{\mu}:=N_{\alpha\mu\beta}g^{\alpha\beta}\;\;, \;\; N^{(3)}_{\mu}:=N_{\mu\alpha\beta}g^{\alpha\beta}
	\eeq 
	respectively. 
\end{mydef}
\begin{mydef} Contracting $N_{\alpha\mu\nu}$ with the Levi-Civita pseudotensor we form the $3$ parity odd combinations as follows
	\beq
	M^{(1)}_{\lambda\alpha\beta}:=N_{\mu\nu\lambda}\varepsilon^{\mu\nu}_{\;\;\;\;\alpha\beta}\;\;, \;\; 	M^{(2)}_{\nu\alpha\beta}:=N_{\mu\nu\lambda}\varepsilon^{\mu\lambda}_{\;\;\;\;\alpha\beta}\;\;, \;\; M^{(3)}_{\mu\alpha\beta}:=N_{\mu\nu\lambda}\varepsilon^{\nu\lambda}_{\;\;\;\;\alpha\beta}
	\eeq
	Note that by construction each of the $M^{(i)}$'s is antisymmetric in its last pair of indices. In addition, we can construct a $4^{th}$ pseudo-trace  as
	\beq
	M^{\alpha}=N^{(4)\alpha}:=\varepsilon^{\alpha\mu\nu\lambda}N_{\mu\nu\lambda}
	\eeq
	Note also that further contractions of the $M^{(i)}$'s  do not give any new traces since $g^{\mu\nu}M^{(1)}_{\mu\nu\alpha}=-g^{\mu\nu}M^{(2)}_{\mu\nu\alpha}=g^{\mu\nu}M^{(3)}_{\mu\nu\alpha}=-M_{\alpha}$ and the rest are either identically vanishing or proportional to the latter.
\end{mydef}

	Let us now touch upon some of the physical aspects of MAG.
	
	\subsection{Canonical and Metrical Energy Momentum and Hypermomentum Tensors}
	In the holonomic\footnote{In an anholonomic description we have three independent fields $g_{ab}, \vartheta^{b}$ and $\Gamma^{a}_{\;\; b}$, that is the tangent space metric, the co-frame and the spin connection. However, the metric field equations are redundant in this case and therefore we have only two independent fields in this case as well (see \cite{hehl1995metric}). } formulation of MAG we are considering here,  one regards the  affine connection $\Gamma^{\lambda}_{\;\;\;\mu\nu}$ and the metric $g_{\mu\nu}$ as independent fundamental fields. Then, the field equations of the Theory are obtained by varying the total action independently with respect the aforementioned fields. Consequently, the variations of the matter part of the action would be the sources of Gravity. Let $\mathcal{L}_{m}$ be the matter Lagrangian of the Theory. As usual we define the energy-momentum tensor of matter by the metric variation of the matter sector, namely 
	\beq
	T^{\alpha\beta}:=+\frac{2}{\sqrt{-g}}\frac{\delta(\sqrt{-g} \mathcal{L}_{M})}{\delta g_{\alpha\beta}}
	\eeq
	Additionally, since we now have an independent affine connection, the variation of the matter part of the action with respect to the latter defines a new tensor called hypermomentum \cite{hehl1976hypermomentum}
	\beq
	\Delta_{\lambda}^{\;\;\;\mu\nu}:= -\frac{2}{\sqrt{-g}}\frac{\delta ( \sqrt{-g} \mathcal{L}_{M})}{\delta \Gamma^{\lambda}_{\;\;\;\mu\nu}}
	\eeq
	Note that the hypermomentum can be split into the three  irreducible pieces of spin, dilation and shear as follows
	\beq
	\Delta_{\mu\nu\alpha}=\tau_{\mu\nu\alpha}+\frac{1}{n}\Delta_{\alpha} g_{\mu\nu}+\hat{\Delta}_{\mu\nu\alpha}
	\eeq
	where $\tau_{\mu\nu\alpha}:=\Delta_{[\mu\nu]\alpha}$ is the spin part, $\Delta_{\alpha}:=\Delta_{\mu\nu\alpha}g^{\mu\nu}$ the dilation (trace) and $\hat{\Delta}_{\mu\nu\alpha}=\Delta_{(\mu\nu)\alpha}-\frac{1}{n}\Delta_{\alpha} g_{\mu\nu}$ the shear (symmetric traceless part).
	The above sources are not totally independent but rather they are subject to the conservation law
	\beq
	\sqrt{-g}(2 \tilde{\nabla}_{\mu}T^{\mu}_{\;\;\alpha}-\Delta^{\lambda\mu\nu}R_{\lambda\mu\nu\alpha})+\hat{\nabla}_{\mu}\hat{\nabla}_{\nu}(\sqrt{-g}\Delta_{\alpha}^{\;\;\mu\nu})+2S_{\mu\alpha}^{\;\;\;\;\lambda}\hat{\nabla}_{\nu}(\sqrt{-g}\Delta_{\lambda}^{\;\;\;\mu\nu})=0\;\;, \;\; \hat{\nabla}_{\mu}:=2 S_{\mu}-\nabla_{\mu} \label{ccc}
	\eeq
	which serves as the generalization of the conservation law of energy-momentum for matter with microstructure  and comes about from the diffeomorphism invariance of the matter part of the action \cite{Iosifidis:2020gth}.

	\section{Full Quadratic Theory}
	
	As we already mentioned the mere inclusion of the Ricci scalar as the gravitational Lagrangian for MAG not only is it problematic due to its projective invariance\footnote{In the sense that it imposes a vanishing dilation current. However in certain instances (see \cite{BeltranJimenez:2019acz,aoki2019scalar}) projective invariance is a key requirement in establishing healthy Theories. }, but it  also is not the only invariant with dimensions $L^{-2}$. Indeed, since now we also have torsion and non-metricity, any quadratic combination of them would be of the same dimension as $R$  and since there is no fundamental principle that excludes them, their  presence on the gravitational action is indispensable. It turns out (see for instance \cite{pagani2015quantum,iosifidis2019scale}) that in any dimension there are $5$ pure on-metricity,  $3$ pure torsion, and $3$ mixed invariants which are quadratic in torsion and non-metricity and parity even. So, in total (and for any dimension) we have the following $11$ parity even quadratic invariants:
	
		\textbf{Pure Non-Metricity }
	\begin{gather}
	A_{1}=Q_{\alpha\mu\nu}Q^{\alpha\mu\nu} \\
	A_{2}=Q_{\alpha\mu\nu}Q^{\mu\nu\alpha} \\
	A_{3}=Q_{\mu}Q^{\mu}
	\\
	A_{4}=q_{\mu}q^{\mu}
	\\
	A_{5}=Q_{\mu}q^{\mu}
	\end{gather}

	\textbf{Pure Torsion }
	\begin{gather}
	B_{1}=S_{\alpha\mu\nu}S^{\alpha\mu\nu} \\
	B_{2}=S_{\alpha\mu\nu}S^{\mu\nu\alpha} \\
	B_{3}=S_{\mu}S^{\mu}
	\end{gather}
	
	\textbf{Mixed}
	\begin{gather}
	C_{1}=Q_{\alpha\mu\nu}S^{\alpha\mu\nu}
	\\
	C_{2}=Q_{\mu}S^{\mu} 
	\\
	C_{3}=q_{\mu}S^{\mu}
	\end{gather}
The above invariants will then supplement the Einstein-Hilbert term (i.e. $R$) into the gravitational action to built the parity even quadratic MAG \cite{hehl1995metric,pagani2015quantum,iosifidis2019scale}. The aforementioned inclusion is fairly general and holds true for any dimension. Now, interestingly enough, in our physical $n=4$ spacetime and with the use of the Levi-Civita pseudo-tensor $\varepsilon^{\mu\nu\alpha\beta}$ additional quadratic (pseudo) scalars can be built. Indeed, in our four-dimensional spacetime we find the $6$ independent pseudoscalars (see \cite{pagani2015quantum,iosifidis2019scale})

\textbf{Parity Odd Invariants}
\begin{gather}
A_{6}=\epsilon^{\alpha\beta\gamma\delta}Q_{\alpha\beta\mu}Q_{\gamma\delta}^{\;\;\;\;\mu} \\
B_{5}=S_{\mu}t^{\mu} \\
B_{6}=\epsilon^{\alpha\beta\gamma\delta}S_{\alpha\beta\mu}S_{\gamma\delta}^{\;\;\;\;\mu}  \\
C_{4}=Q_{\mu}t^{\mu}  \\
C_{5}=q^{\mu}t_{\mu}  \\
C_{6}=\epsilon^{\alpha\beta\gamma\delta}Q_{\alpha\beta\mu}S_{\gamma\delta}^{\;\;\;\;\mu}
\end{gather}
Note that other quadratic combinations could be considered, such as $t_{\mu}t^{\mu}$ for instance. However, these can be shown to be related to the ones written above and so there are exactly $6$ parity odd quadratic invariants in torsion and non-metricity.

	Now, is this all there is to it?  Well, our initial intention was to built a gravitational action consisting of scalars that  are linear in the Riemann tensor and quadratic in torsion and non-metricity since all of these will be of the same dimension. The above exposition has exhausted the possibilities for the quadratic terms so the question is how many scalars can we built out of the Riemann tensor? Of course, the scalar that is constructed by (and is linear in) the Riemann tensor is the Ricci scalar $R$. For a symmetric connection this is the one and only possibility. Interestingly, for a general affine connection, as the one we are considering here, in $n=4$ and given the existence of the Levi-Civita pseudotensor one can include the totally antisymmetric part of the Riemann tensor, which is given by the pseudoscalar\footnote{Here $\varepsilon^{\mu\nu\alpha\beta}$ is the Levi-Civita pesudo-tensor which is related to the Levi-Civita symbol $\epsilon^{\mu\nu\rho\sigma}$ through $	\varepsilon^{\mu\nu\rho\sigma}:=\frac{\epsilon^{\mu\nu\rho\sigma}}{\sqrt{-g}}$.} 
	\beq
	\varepsilon^{\mu\nu\alpha\beta}R_{\mu\nu\alpha\beta} \label{eR}
	\eeq
	and which is of purely non-Riemannian nature\footnote{More precisely, the emergence of this term has to do with torsion a is clearly demonstrated in an identity we give below. In other words, for vanishing torsion $\varepsilon^{\mu\nu\alpha\beta}R_{\mu\nu\alpha\beta}$ vanishes as well.}. The above totally antisymmetric part is sometimes referred to as as the Hojman, or most commonly called the Holst term\footnote{This term was first studied by Hojman et al in \cite{PhysRevD.22.1915} and then later by Holst \cite{holst1996barbero}. Its inclusion in Gravity Theories has regained a lot of attention recently and has been studied in various contexts (see \cite{Iosifidis:2020dck,Bombacigno:2021bpk,Boudet:2020eyr,Obukhov:2020zal,Bombacigno:2019nua,laangvik2021higgs}).}.  
		Therefore, the most general gravitational MAG action (in $n=4$) that is linear in curvature and quadratic in torsion and non-metricity is given by
	\begin{gather}
	S
	=\frac{1}{2 \kappa}\int d^{4}x \sqrt{-g} \Big[ R+\lambda   \varepsilon^{\alpha\beta\gamma\delta}R_{\alpha\beta\gamma\delta}+
	b_{1}S_{\alpha\mu\nu}S^{\alpha\mu\nu} +
	b_{2}S_{\alpha\mu\nu}S^{\mu\nu\alpha} +
	b_{3}S_{\mu}S^{\mu} \nonumber \\
	a_{1}Q_{\alpha\mu\nu}Q^{\alpha\mu\nu} +
	a_{2}Q_{\alpha\mu\nu}Q^{\mu\nu\alpha} +
	a_{3}Q_{\mu}Q^{\mu}+
	a_{4}q_{\mu}q^{\mu}+
	a_{5}Q_{\mu}q^{\mu} \nonumber \\
	+c_{1}Q_{\alpha\mu\nu}S^{\alpha\mu\nu}+
	c_{2}Q_{\mu}S^{\mu} +
	c_{3}q_{\mu}S^{\mu} \nonumber \\
	+a_{6}\varepsilon^{\alpha\beta\gamma\delta}Q_{\alpha\beta\mu}Q_{\gamma\delta\;\;\;\;}^{\mu}+b_{5}S_{\mu}t^{\mu}+b_{6}\varepsilon^{\alpha\beta\gamma\delta}S_{\alpha\beta\mu}S_{\gamma\delta}^{\;\;\;\;\mu} \nonumber \\
	c_{4}Q_{\mu}t^{\mu}+c_{5}q^{\mu}t_{\mu}+c_{6}\varepsilon^{\alpha\beta\gamma\delta}Q_{\alpha\beta\mu}S_{\gamma\delta}^{\;\;\;\;\mu}
	\Big]  \label{S}
	\end{gather}
	  However, it is important to stress out that there exists the identity (for the proof see for instance \cite{iosifidis2019metric})
		\beq
		\varepsilon^{\mu\nu\alpha\beta}R_{\mu\nu\alpha\beta}=2 \tilde{\nabla}_{\alpha}\tilde{S}^{\alpha}+2 \varepsilon^{\mu\nu\alpha\beta}S_{\alpha\beta}^{\;\;\;\;\lambda}( Q_{\mu\nu\lambda}+S_{\mu\nu\lambda}) \,,
		\eeq
	where we have used the decomposition ($\ref{N}$) for the Riemann tensor, the fact that $\varepsilon^{\mu\nu\alpha\beta}\tilde{R}_{\mu\nu\alpha\beta}=0$. From the above identity we see that the first term will result in a total derivative term which will not contribute to the field equations. Furthermore, we see that the second and third terms are already included in the parity odd term classification we outlined earlier. More precisely, these terms are respectively equal (up to a factor of $2$) to the invariants $C_{6}$ and $B_{6}$ we presented above. Therefore, the inclusion of ($\ref{eR}$) into the Gravitational action would simply amount to a shift on the coefficients $c_{6}$ and $b_{6}$ of the terms $\varepsilon^{\alpha\beta\gamma\delta}Q_{\alpha\beta\mu}S_{\gamma\delta}^{\;\;\;\;\mu}$ and $\varepsilon^{\alpha\beta\gamma\delta}S_{\alpha\beta\mu}S_{\gamma\delta}^{\;\;\;\;\mu}$ respectively. Therefore the inclusion of $(\ref{eR})$ is redundant and we can always safely set $\lambda=0$ in ($\ref{S}$) without altering the physics of the Theory at classical level.

	\section{The full Quadratic MAG Theory}
	As we already mentioned the main intention of this work is to analyze the full $17$-parameter quadratic MAG Theory including all possible, parity even and parity odd,  quadratic terms in torsion and non-metricity in addition to the Einstein-Hilbert term. Even though of great importance is to answer what exactly   the vacuum Theory corresponds to, even more interesting and challenging is to add matter and study the full Theory in the presence of sources. Of course, having done so we can always turn off the sources afterwards and focus on the vacuum case. As we shall show, regardless of the complexity of the $17$ parameter full quadratic Theory, we will be able to answer to both of the aforementioned questions. Our $4$ dimensional $17$ parameter full quadratic Theory reads
	\begin{gather}
	S[g, \Gamma, \Phi]
	=\frac{1}{2 \kappa}\int d^{4}x \sqrt{-g} \Big[  R+ 
	b_{1}S_{\alpha\mu\nu}S^{\alpha\mu\nu} +
	b_{2}S_{\alpha\mu\nu}S^{\mu\nu\alpha} +
	b_{3}S_{\mu}S^{\mu} \nonumber \\
	a_{1}Q_{\alpha\mu\nu}Q^{\alpha\mu\nu} +
	a_{2}Q_{\alpha\mu\nu}Q^{\mu\nu\alpha} +
	a_{3}Q_{\mu}Q^{\mu}+
	a_{4}q_{\mu}q^{\mu}+
	a_{5}Q_{\mu}q^{\mu} \nonumber \\
	+c_{1}Q_{\alpha\mu\nu}S^{\alpha\mu\nu}+
	c_{2}Q_{\mu}S^{\mu} +
	c_{3}q_{\mu}S^{\mu} \nonumber \\
	+a_{6}\varepsilon^{\alpha\beta\gamma\delta}Q_{\alpha\beta\mu}Q_{\gamma\delta}^{\;\;\;\;\mu}+b_{5}S_{\mu}t^{\mu}+b_{6}\varepsilon^{\alpha\beta\gamma\delta}S_{\alpha\beta\mu}S_{\gamma\delta}^{\;\;\;\;\mu} \nonumber \\
	c_{4}Q_{\mu}t^{\mu}+c_{5}q^{\mu}t_{\mu}+c_{6}\varepsilon^{\alpha\beta\gamma\delta}Q_{\alpha\beta\mu}S_{\gamma\delta}^{\;\;\;\;\mu}
	\Big] +S_{M}[g, \Gamma, \Phi] \label{action}
	\end{gather}
where we have set $\lambda=0$ since, as demonstrated previously, this term is already included in the quadratic terms.	In the above, the first, second and third lines contain (apart from the usual Einstein-Hilbert term) the pure torsion, pure non-metricity and mixed quadratic parity even parts while the last two lines contain all the parity odd terms. Let us highlight that the matter action we are considering here can (and will) in general depend on the connection which will result in a non-vanishing hypermomentum. Varying the above action independently with respect to the metric and the affine connection, after some long calculations we obtain the field equations
	\begin{gather}
R_{(\mu\nu)}-\frac{R}{2}g_{\mu\nu}-\frac{\mathcal{L}^{(2)}_{even}}{2}g_{\mu\nu}
	-\frac{1}{\sqrt{-g}}(\nabla_{\alpha}-2 S_{\alpha})\Big[ \sqrt{-g}\Big( c_{1}S^{\alpha}_{\;\;(\mu\nu)}+c_{2}g_{\mu\nu}S^{\alpha}+c_{3}\delta^{\alpha}_{(\mu}S_{\nu)}\Big)\Big]	\nonumber \\
	-\frac{1}{\sqrt{-g}}(\nabla_{\alpha}-2 S_{\alpha})\Big[ \sqrt{-g}\Big(2 a_{1}Q^{\alpha}_{\;\;\mu\nu}+2 a_{2}Q_{(\mu\nu)}^{\;\;\;\;\alpha}+(2 a_{3}Q^{\alpha}+a_{5}q^{\alpha})g_{\mu\nu}+(2 a_{4}q_{(\mu} + a_{5}Q_{(\mu})\delta^{\alpha}_{\nu)}\Big)\Big]	\nonumber \\
	+	a_{1}(Q_{\mu\alpha\beta}Q_{\nu}^{\;\;\alpha\beta}-2 Q_{\alpha\beta\mu}Q^{\alpha\beta}_{\;\;\;\;\nu})-a_{2}Q_{\alpha\beta(\mu}Q^{\beta\alpha}_{\;\;\;\;\nu)}
		+a_{3}(Q_{\mu}Q_{\nu}-2 Q^{\alpha}Q_{\alpha\mu\nu})-a_{4}q_{\mu}q_{\nu}-a_{5}q^{\alpha}Q_{\alpha\mu\nu}	\nonumber \\
	+b_{1}(2S_{\nu\alpha\beta}S_{\mu}^{\;\;\;\alpha\beta}-S_{\alpha\beta\mu}S^{\alpha\beta}_{\;\;\;\;\nu})-b_{2}S_{(\mu}^{\;\;\;\beta\alpha}S_{\nu)\alpha\beta}+b_{3}S_{\mu}S_{\nu} 	\nonumber \\
	+c_{1}(Q_{(\mu}^{\;\;\;\alpha\beta}S_{\nu)\alpha\beta}-S_{\alpha\beta(\mu}Q^{\alpha\beta}_{\;\;\;\;\nu)})+c_{2}(S_{(\mu}Q_{\nu)}-S^{\alpha}Q_{\alpha\mu\nu})\nonumber \\
		-b_{6} \varepsilon^{\alpha\beta\gamma\delta} S_{\alpha\beta\nu}S_{\gamma\delta\mu}+ a_{6}\varepsilon^{\alpha\beta\gamma\delta}Q_{\alpha\beta\mu}Q_{\gamma\delta\nu} \nonumber \\
		- g_{\beta(\mu}g_{\nu)\lambda}\frac{1}{\sqrt{-g}}(\nabla_{\alpha}- 2 S_{\alpha})\Big(\sqrt{-g}  \varepsilon^{\alpha\beta\gamma\delta} ( 2 a_{6} Q_{\gamma\delta}^{\;\;\;\;\lambda}+c_{6} S_{\gamma\delta}^{\;\;\;\;\lambda}) \Big)\nonumber \\
		-S_{\alpha\beta(\mu}\varepsilon^{\lambda\alpha\beta}_{\;\;\;\;\;\;\nu)}(b_{5} S_{\lambda}+c_{4}Q_{\lambda}+c_{5} q_{\lambda})
		-c_{4}g_{\mu\nu}\Big(\partial_{\lambda}(\sqrt{-g}t^{\lambda})\Big)\nonumber \\
		-c_{5}g_{\alpha(\mu}g_{\nu)\beta}\frac{1}{\sqrt{-g}}\nabla_{\lambda}(\sqrt{-g}t^{\alpha}g^{\lambda\beta})+ c_{5}2 S_{(\mu}t_{\nu)}+c_{5}t^{\alpha}Q_{(\mu\nu)\alpha} 
		=\kappa T_{\mu\nu} \label{metricf}
\end{gather}

	\textbf{$\Gamma$-Variation}
	
	\begin{gather}
\left( \frac{Q_{\lambda}}{2}+2 S_{\lambda}\right)g^{\mu\nu}-Q_{\lambda}^{\;\;\mu\nu}-2 S_{\lambda}^{\;\;\mu\nu}+\left( q^{\mu} -\frac{Q^{\mu}}{2}-2 S^{\mu}\right)\delta_{\lambda}^{\nu}+4 a_{1}Q^{\nu\mu}_{\;\;\;\;\lambda}+2 a_{2}(Q^{\mu\nu}_{\;\;\;\;\lambda}+Q_{\lambda}^{\;\;\;\mu\nu})+2 b_{1}S^{\mu\nu}_{\;\;\;\;\lambda} \nonumber \\
+2 b_{2}S_{\lambda}^{\;\;\;[\mu\nu]}+c_{1}\Big( S^{\nu\mu}_{\;\;\;\;\lambda}-S_{\lambda}^{\;\;\;\nu\mu}+Q^{[\mu\nu]}_{\;\;\;\;\;\lambda}\Big)+\delta_{\lambda}^{\mu}\Big( 4 a_{3}Q^{\nu}+2 a_{5}q^{\nu}+2 c_{2}S^{\nu}\Big)+\delta_{\lambda}^{\nu}\Big(  a_{5}Q^{\mu}+2 a_{4}q^{\mu}+ c_{3}S^{\mu}\Big) \nonumber \\
+g^{\mu\nu}\Big(a_{5} Q_{\lambda}+2 a_{4}q_{\lambda}+c_{3}S_{\lambda} \Big)+\Big( c_{2} Q^{[\mu}+ c_{3}q^{[\mu}+2 b_{3}S^{[\mu}\Big) \delta^{\nu]}_{\lambda} \nonumber \\
+(-2 a_{6}+c_{6})\varepsilon^{\mu\nu\alpha\beta}Q_{\alpha\beta\lambda}+(2 b_{6}-c_{6})\varepsilon^{\mu\nu\alpha\beta}S_{\alpha\beta\lambda}-2 a_{6}\varepsilon_{\lambda}^{\;\;\nu\alpha\beta}Q_{\alpha\beta}^{\;\;\;\;\mu}-c_{6}\varepsilon_{\lambda}^{\;\;\nu\alpha\beta}S_{\alpha\beta}^{\;\;\;\;\mu} \nonumber \\
+\varepsilon^{\alpha\mu\nu}_{\;\;\;\;\;\;\lambda}(b_{5}S_{\alpha}+c_{4}Q_{\alpha}+c_{5}q_{\alpha})+\Big( \frac{b_{5}}{2}+c_{5}\Big) t^{\mu}\delta_{\lambda}^{\nu}+\Big( -\frac{b_{5}}{2}+2c_{4}\Big) t^{\nu}\delta_{\lambda}^{\mu}+c_{5}g^{\mu\nu}t_{\lambda} =\kappa \Delta_{\lambda}^{\;\;\;\mu\nu} \label{Gfieldeqs}
\end{gather}
We see that both of the above sets of field equations look somewhat complicated. However, as we shall show, using a recent result (see \cite{iosifidis2021solving}), quite remarkably, we will be able to solve completely (and exactly) for the affine connection and subsequently recast the metric field equations as an effective Einstein's GR with modified/extended source terms that are associated to the microstructure of matter. We do so in what follows.
	
	\section{Solving for the Affine Connection}
	
	We shall now undertake the challenge of solving for the affine connection. Even though the situation seems a little messy, using a certain Theorem, we will provide the unique and exact solution of ($\ref{Gfieldeqs}$) given some general non-degeneracy conditions. Of paramount importance in arriving at such a solution is the following result which we give here without proof. The detailed proof can be found in \cite{iosifidis2021solving}.

\begin{theo}
In a $4$-dimensional space of signature $s$, consider the $30$ parameter tensor equation
\begin{gather}
a_{1}N_{\alpha\mu\nu}+a_{2}N_{\nu\alpha\mu}+a_{3}N_{\mu\nu\alpha}+a_{4}N_{\alpha\nu\mu}+a_{5}N_{\nu\mu\alpha}+a_{6}N_{\mu\alpha\nu}+\sum_{i=1}^{3}\Big( a_{7 i}N^{(i)}_{\mu}g_{\alpha\nu}+a_{8 i}N^{(i)}_{\nu}g_{\alpha\mu}+a_{9 i}N^{(i)}_{\alpha}g_{\mu\nu} \Big) \nonumber \\
+b_{11}M^{(1)}_{\alpha\mu\nu}+b_{12}M^{(1)}_{\nu\alpha\mu}+b_{13}M^{(1)}_{\mu\nu\alpha}+b_{21}M^{(2)}_{\alpha\mu\nu}+b_{22}M^{(2)}_{\nu\alpha\mu}+b_{23}M^{(2)}_{\mu\nu\alpha}+b_{31}M^{(3)}_{\alpha\mu\nu}+b_{32}M^{(3)}_{\nu\alpha\mu}+b_{33}M^{(3)}_{\mu\nu\alpha} \nonumber \\
+\varepsilon_{\rho\alpha\mu\nu}\Big( b_{1}N^{(1)\rho}+ b_{2}N^{(2)\rho}+ b_{3}N^{(3)\rho}\Big)	+c_{1}M_{\mu}g_{\alpha\nu}+c_{2}M_{\nu}g_{\alpha\mu}+c_{3}M_{\alpha}g_{\mu\nu}
=B_{\alpha\mu\nu} \label{eq1}
\end{gather}
where  $a_{i}$, $a_{ji}$ $i=1,2,...,6$, $j=7,8,9$ are scalars, $b_{kl}$, $c_{m}$ are pseudo-scalars,  $B_{\alpha\mu\nu}$   is a given (known) tensor and $N_{\alpha\mu\nu}$ are the components of the unknown tensor\footnote{Of course the result holds true even when $N_{\alpha\mu\nu}$ are the components of a tensor density instead or even of a connection given that $B_{\alpha\mu\nu}$ are also  of the same kind. } $N$. Define the matrices
\setcounter{MaxMatrixCols}{20}
\begin{equation}
A := 
\begin{pmatrix}
\alpha_{11} & \alpha_{12} & \alpha_{13} & \alpha_{14} & \alpha_{15} & \alpha_{16} & \alpha_{17} & \alpha_{18} & \alpha_{19} & \alpha_{1,10}&\alpha_{1,11} & \alpha_{1,12} & \alpha_{1,13} & \alpha_{1,14} & \alpha_{1,15}  \\
\alpha_{21} & \alpha_{22} & \alpha_{23} & \alpha_{24} & \alpha_{25} & \alpha_{26} & \alpha_{27} & \alpha_{28} & \alpha_{29} & \alpha_{2,10} & \alpha_{2,11} & \alpha_{1,12} & \alpha_{2,13} & \alpha_{2,14} & \alpha_{2,15} \\
\alpha_{31} & \alpha_{32} & \alpha_{33} & \alpha_{34} & \alpha_{35} & \alpha_{36} & \alpha_{37} & \alpha_{38} & \alpha_{39} & \alpha_{3,10} & \alpha_{3,11} & \alpha_{1,12} & \alpha_{1,13} & \alpha_{1,14} & \alpha_{1,15}  \\
\alpha_{41} & \alpha_{42} & \alpha_{43} & \alpha_{44} & \alpha_{45} & \alpha_{46} & \alpha_{47} & \alpha_{48} & \alpha_{49} & \alpha_{4,10} & \alpha_{4,11} &  \alpha_{4,12} & \alpha_{4,13} & \alpha_{4,14} & \alpha_{4,15} \\
\alpha_{51} & \alpha_{52} & \alpha_{53} & \alpha_{54} & \alpha_{55} & \alpha_{56} & \alpha_{57} & \alpha_{58} & \alpha_{59} & \alpha_{5,10} & \alpha_{5,11} & \alpha_{5,12} & \alpha_{5,13} & \alpha_{5,14} & \alpha_{5,15} \\
\alpha_{61} & \alpha_{62} & \alpha_{63} & \alpha_{64} & \alpha_{65} & \alpha_{66} & \alpha_{67} & \alpha_{68} & \alpha_{69} & \alpha_{6,10} & \alpha_{6,11} & \alpha_{6,12} & \alpha_{6,13} & \alpha_{6,14} & \alpha_{6,15}  \\
\alpha_{71} & \alpha_{72} & \alpha_{73} & \alpha_{74} & \alpha_{75} & \alpha_{76} & \alpha_{77} & \alpha_{78} & \alpha_{79} & \alpha_{7,10} & \alpha_{7,11} & \alpha_{7,12} & \alpha_{7,13} & \alpha_{7,14} & \alpha_{7,15}  \\
\alpha_{81} & \alpha_{82} & \alpha_{83} & \alpha_{84} & \alpha_{85} & \alpha_{86} & \alpha_{87} & \alpha_{88} & \alpha_{89} & \alpha_{8,10} & \alpha_{8,11} & \alpha_{8,12} & \alpha_{8,13} & \alpha_{8,14} & \alpha_{8,15} \\
\alpha_{91} & \alpha_{92} & \alpha_{93} & \alpha_{94} & \alpha_{95} & \alpha_{96} & \alpha_{97} & \alpha_{98} & \alpha_{99} & \alpha_{9,10} & \alpha_{9,11} &  \alpha_{9,12} &\alpha_{9,13} & \alpha_{9,14} & \alpha_{9,15}  \\
\alpha_{10,1} & \alpha_{10,2} & \alpha_{10,3} & \alpha_{10,4} & \alpha_{10,5} & \alpha_{10,6} & \alpha_{10,7} & \alpha_{10,8} & \alpha_{10,9} & \alpha_{10,10} & \alpha_{10,11} & \alpha_{10,12} & \alpha_{10,13} & \alpha_{10,14} & \alpha_{10,15} \\
\alpha_{11,1} & \alpha_{11,2} & \alpha_{11,3} & \alpha_{11,4} & \alpha_{11,5} & \alpha_{11,6} & \alpha_{11,7} & \alpha_{11,8} & \alpha_{11,9} & \alpha_{11,10} & \alpha_{11,11} & \alpha_{11,12} & \alpha_{11,13} & \alpha_{11,14} & \alpha_{11,15} \\
\alpha_{12,1} & \alpha_{12,2} & \alpha_{12,3} & \alpha_{12,4} & \alpha_{12,5} & \alpha_{12,6} & \alpha_{12,7} & \alpha_{12,8} & \alpha_{12,9} & \alpha_{12,10} & \alpha_{12,11} & \alpha_{12,12} & \alpha_{12,13} & \alpha_{12,14} & \alpha_{12,15}  \\
\alpha_{13,1} & \alpha_{13,2} & \alpha_{13,3} & \alpha_{14,4} & \alpha_{14,5} & \alpha_{14,6} & \alpha_{14,7} & \alpha_{14,8} & \alpha_{14,9} & \alpha_{1,10} & \alpha_{1,11} & \alpha_{1,12} & \alpha_{1,13} & \alpha_{1,14} & \alpha_{1,15}  \\
\alpha_{11} & \alpha_{12} & \alpha_{13} & \alpha_{14} & \alpha_{15} & \alpha_{16} & \alpha_{17} & \alpha_{18} & \alpha_{19} & \alpha_{1,10} & \alpha_{14,11} & \alpha_{14,12} & \alpha_{14,13} & \alpha_{14,14} & \alpha_{14,15}  \\
\alpha_{15,1} & \alpha_{15,2} & \alpha_{15,3} & \alpha_{15,4} & \alpha_{15,5} & \alpha_{15,6} & \alpha_{15,7} & \alpha_{15,8} & \alpha_{15,9} & \alpha_{15,10} & \alpha_{15,11} &  \alpha_{15,12} &\alpha_{15,13} & \alpha_{15,14} & \alpha_{15,15}  \\
\end{pmatrix} \label{A}
\end{equation}
and 
\begin{equation}
\Gamma := 
\begin{pmatrix}
\gamma_{11} & \gamma_{12} & \gamma_{13} & \gamma_{14}  \\
\gamma_{21} & \gamma_{22} & \gamma_{23} & \gamma_{24}  \\
\gamma_{31} & \gamma_{32} & \gamma_{33} & \gamma_{34}  \\
\gamma_{41} & \gamma_{42} & \gamma_{43} & \gamma_{44}  \\
\end{pmatrix} \label{G}
\end{equation}		
where  the $\alpha_{ij}$ and  $\gamma_{ij}$'s are some linear combinations of the original $a_{i}$, $a_{ji}$, $b_{j}$ and $c_{k}$'s (see appendix). Then given that both of these matrices are non-singular, that is if	
\beq
det(A) \neq 0 \;\; and \;\; det(\Gamma) \neq 0
\eeq
holds true, then the general and unique solution of ($\ref{eq1}$) reads
\begin{gather}
N_{\alpha\mu\nu}=\tilde{\alpha}_{11}\hat{B}_{\alpha\mu\nu}+\tilde{\alpha}_{12}\hat{B}_{\nu\alpha\mu}+\tilde{\alpha}_{13} \hat{B}_{\mu\nu\alpha}+\tilde{\alpha}_{14} \hat{B}_{\alpha\nu\mu}+\tilde{\alpha}_{15} \hat{B}_{\nu\mu\alpha} +\tilde{\alpha}_{16} \hat{B}_{\mu\alpha\nu}+\tilde{\alpha}_{17}\breve{B}_{\alpha\mu\nu}+\tilde{\alpha}_{18} \breve{B}_{\mu\nu\alpha} + \tilde{\alpha}_{19}\breve{B}_{\nu\alpha\mu} \nonumber \\
+\tilde{\alpha}_{1,10}\bar{B}_{\alpha\mu\nu}+\tilde{\alpha}_{1,11}\bar{B}_{\mu\nu\alpha}+\tilde{\alpha}_{1,12}\bar{B}_{\nu\alpha\mu}+\tilde{\alpha}_{1,13}\mathring{B}_{\alpha\mu\nu}+\tilde{\alpha}_{1,14}\mathring{B}_{\mu\nu\alpha}+\tilde{\alpha}_{1,15}\mathring{B}_{\nu\alpha\mu} \label{MasterN}
\end{gather}
where the $\tilde{\alpha}_{1i}$ $  's$ are the first row elements of the inverse matrix $A^{-1}$ and 
\beq
\hat{B}_{\alpha\mu\nu} =B_{\alpha\mu\nu} -\sum_{i=1}^{4}\sum_{j=1}^{4}\Big( a_{7i}\tilde{\gamma}_{ij}B^{(j)}_{\mu}g_{\alpha\nu}+a_{8i}\tilde{\gamma}_{ij}B^{(j)}_{\nu}g_{\alpha\mu}+ a_{9i}\tilde{\gamma}_{ij}B^{(j)}_{\alpha}g_{\mu\nu} \Big)-\varepsilon^{\rho}_{\;\;\alpha\mu\nu}\sum_{i=1}^{3}\sum_{j=1}^{4}b_{i}\tilde{\gamma}_{ij}B^{(j)}_{\rho} \label{B1}
\eeq
\beq
\breve{B}_{\alpha\mu\nu}=	\varepsilon^{\beta\gamma}_{\;\;\;\;\alpha\mu}\hat{B}_{\beta\gamma\nu}-2(-1)^{s}\sum_{j=1}^{4}\Big[ (b_{21}+b_{23}+b_{31}+b_{33})\tilde{\gamma}_{1j}+(b_{11}+b_{13}-b_{31}-b_{33})\tilde{\gamma}_{2j}-(b_{11}+b_{13}+b_{21}+b_{23})\tilde{\gamma}_{3j}\Big]B^{(j)}_{[\alpha}g_{\mu]\nu} \label{B2}
\eeq
\beq
\bar{B}_{\alpha\mu\nu}:=	\varepsilon^{\beta\gamma}_{\;\;\;\;\alpha\nu}\hat{B}_{\beta\mu\gamma}-2(-1)^{s}\sum_{j=1}^{4}\Big[ (b_{21}+b_{22}+b_{31}+b_{32})\tilde{\gamma}_{1j}+(b_{11}+b_{12}-b_{31}-b_{32})\tilde{\gamma}_{2j}-(b_{11}+b_{12}+b_{21}+b_{22})\tilde{\gamma}_{3j}\Big]B^{(j)}_{[\alpha}g_{\mu]\nu} \label{B3}
\eeq
\beq
\mathring{B}_{\alpha\mu\nu}:=	\varepsilon^{\beta\gamma}_{\;\;\;\;\mu\nu}	\mathring{B}_{\alpha\beta\gamma}-2(-1)^{s+1}\sum_{j=1}^{4}\Big[ (b_{22}+b_{23}+b_{32}+b_{33})\tilde{\gamma}_{1j}+(b_{12}+b_{13}-b_{32}-b_{33})\tilde{\gamma}_{2j}-(b_{22}+b_{23}+b_{12}+b_{13})\tilde{\gamma}_{3j}\Big]B^{(j)}_{[\mu}g_{\nu]\alpha} \label{B4}
\eeq
$\tilde{\gamma}_{ij}'s$ being the elements of the inverse matrix $\Gamma^{-1}$.
\end{theo}
From the above, it also immediately follows that 
 when $B_{\alpha}=0$ and $det(A)\neq 0$ along with $det(\Gamma)\neq 0$  we have that the unique solution of $(\ref{eq1})$ is always $N_{\alpha\mu\nu}=0$. More precisely, we have the following \cite{iosifidis2021solving}
\begin{corollary}
	If $B_{\alpha\mu\nu}=0$ and both matrices $A$ and $\Gamma$ are non-singular, then the unique solution of $(\ref{eq1})$ is $N_{\alpha\mu\nu}=0$.
\end{corollary}

With the above result,  despite  the seemingly complicated form of the connection field equations ($\ref{Gfieldeqs}$), as we shall show below the exact form of the affine connection in therms of the hypermomentum can be readily obtained.

The crucial point now is to use equations ($\ref{QNSN}$) and also their  contractions (see appendix) in order eliminate torsion and non-metricity from ($\ref{Gfieldeqs}$) and  express everything in terms of the distortion tensor and its associated vectors. After some lengthy but rather straightforward algebra we finally arrive at

		\begin{gather}
	( 4a_{1}+b_{1}-c_{1})N_{\alpha\mu\nu}+\Big( -1+2 a_{2}+\frac{c_{1}+b_{2}}{2}\Big)N_{\nu\alpha\mu}+\Big( -1+2 a_{2}+\frac{c_{1}+b_{2}}{2}\Big)N_{\mu\nu\alpha}\nonumber \\
	+(2 a_{2}-b_{1}+c_{1})N_{\alpha\nu\mu}+\Big( 2 a_{2}-\frac{b_{2}}{2}\Big)
	N_{\nu\mu\alpha}+\Big( 4 a_{1}-\frac{b_{2}}{2}-c_{1}\Big) N_{\mu\alpha\nu} \nonumber \\
	+\left(2 a_{5}+c_{2}-\frac{b_{3}+c_{3}}{2}\right)g_{\nu\alpha} N^{(1)}_{\mu}+\left(2 a_{4}+\frac{b_{3}}{2}+c_{3}\right)g_{\nu\alpha} N^{(2)}_{\mu}+\left( 1+ 2a_{4}+\frac{c_{3}}{2}\right) g_{\nu\alpha}  N^{(3)}_{\mu} \nonumber \\
	+\left( 8 a_{3}- 2 c_{2}+\frac{b_{3}}{2}\right) g_{\mu\alpha}N^{(1)}_{\nu}+\left( 2 a_{5}+c_{2}-\frac{c_{3}+b_{3}}{2}\right) g_{\mu\alpha}N^{(2)}_{\mu}+\left( 2 a_{5}-\frac{c_{3}}{2}\right) g_{\mu\alpha}N^{(3)}_{\nu} \nonumber \\
	+\left( 2 a_{5}-\frac{c_{3}}{2}\right)g_{\mu\nu} N^{(1)}_{\alpha}+\left( 2 a_{4}+\frac{c_{3}}{2}+1\right) g_{\mu\nu}N^{(2)}_{\alpha}+2 a_{4}g_{\mu\nu}N^{(3)}_{\alpha} \nonumber \\
	-(-2 a_{6}+c_{6})  \varepsilon^{\mu\nu\alpha\beta}N_{\alpha\lambda\beta}-(-2 a_{6}+c_{6})\varepsilon^{\mu\nu\alpha\beta}N_{\lambda\alpha\beta}                         +(2 b_{6}-c_{6})\varepsilon^{\mu\nu\alpha\beta}N_{\alpha\beta\lambda} \nonumber \\
	+2 a_{6}\varepsilon_{\lambda}^{\;\;\nu\alpha\beta}\Big( N_{\alpha\;\;\;\;\beta}^{\;\;\mu}+N^{\mu}_{\;\;\alpha\beta} \Big)-c_{6}\varepsilon_{\lambda}^{\;\;\nu\alpha\beta}N^{\mu}_{\;\;\alpha\beta}	+\varepsilon^{\alpha\mu\nu}_{\;\;\;\;\;\;\lambda} \Big( 2 c_{4}-\frac{b_{5}}{2}\Big) N^{(1)}_{\alpha} \nonumber \\
	+\Big(  c_{5}+\frac{b_{5}}{2}\Big) \varepsilon^{\alpha\mu\nu}_{\;\;\;\;\;\;\lambda} N^{(2)}_{\alpha}+ c_{5}\varepsilon^{\alpha\mu\nu}_{\;\;\;\;\;\;\lambda}N^{(3)}_{\alpha}+\Big( \frac{b_{5}}{2}+c_{5}\Big) M^{\mu}\delta_{\lambda}^{\nu}+\Big( -\frac{b_{5}}{2}+2c_{4}\Big) M^{\nu}\delta_{\lambda}^{\mu}+c_{5}g^{\mu\nu}M_{\lambda}
	=\kappa \Delta_{\alpha\mu\nu} \label{Neq}
	\end{gather}
	Now we see that the latter equation falls in the case of $Theorem-1$ with the obvious identifications among the parameters $\alpha_{i}$ and $a_{i}, b_{j}, c_{k}$ etc. Therefore, in applying $Theorem-1$ to the latter and setting $B_{\alpha\mu\nu}=\kappa \Delta_{\alpha\mu\nu}$ we solve for the distortion
	\begin{gather}
	N_{\alpha\mu\nu}=\kappa\tilde{\alpha}_{11}\hat{\Delta}_{\alpha\mu\nu}+\kappa\tilde{\alpha}_{12}\hat{\Delta}_{\nu\alpha\mu}+\kappa\tilde{\alpha}_{13} \hat{\Delta}_{\mu\nu\alpha}+\kappa\tilde{\alpha}_{14} \hat{\Delta}_{\alpha\nu\mu}+\kappa\tilde{\alpha}_{15} \hat{\Delta}_{\nu\mu\alpha} +\kappa\tilde{\alpha}_{16} \hat{\Delta}_{\mu\alpha\nu}+\kappa\tilde{\alpha}_{17}\breve{\Delta}_{\alpha\mu\nu}
	+\kappa\tilde{\alpha}_{18} \breve{\Delta}_{\mu\nu\alpha} \nonumber \\ + \kappa\tilde{\alpha}_{19}\breve{\Delta}_{\nu\alpha\mu} 
	+\kappa\tilde{\alpha}_{1,10}\bar{\Delta}_{\alpha\mu\nu}+\kappa\tilde{\alpha}_{1,11}\bar{\Delta}_{\mu\nu\alpha}+\kappa\tilde{\alpha}_{1,12}\bar{\Delta}_{\nu\alpha\mu}+\kappa\tilde{\alpha}_{1,13}\mathring{\Delta}_{\alpha\mu\nu}+\kappa\tilde{\alpha}_{1,14}\mathring{\Delta}_{\mu\nu\alpha}+\kappa\tilde{\alpha}_{1,15}\mathring{\Delta}_{\nu\alpha\mu} \label{MasterNB}
	\end{gather}
	The latter is the exact form of the distortion tensor for the full quadratic Theory. Substituting this back at the decomposition ($\ref{N}$) we obtain the exact and unique solution for the affine connection. We may then readily compute the associated forms of spacetime torsion and non-metricity which we give below.
	
	\subsection{Expressions for Torsion and Non-metricity}
	
	Using the solution ($\ref{MasterNB}$) and the relations ($\ref{QNSN}$) we easily derive the forms of torsion and non-metricity which read\footnote{Here vertical bars $||$ around an index indicate that the latter has been left out of the (anti-)symmetrization process.}
	\begin{gather}
	S_{\mu\nu\alpha}=\kappa \Big[ (\tilde{\alpha}_{11}-\tilde{\alpha}_{14})\hat{\Delta}_{\alpha[\mu\nu]}+\tilde{\alpha}_{17} \breve{\Delta}_{\alpha[\mu\nu]}+\tilde{\alpha}_{1,10}\bar{\Delta}_{\alpha[\mu\nu]}\Big] \nonumber \\
	+\kappa \Big[ (\tilde{\alpha}_{16}-\tilde{\alpha}_{12})\hat{\Delta}_{[\mu|\alpha|\nu]}-\tilde{\alpha}_{19} \breve{\Delta}_{[\mu|\alpha|\nu]}-\tilde{\alpha}_{1,15}\mathring{\Delta}_{[\mu|\alpha|\nu]}\Big]\nonumber \\
		+\kappa \Big[ (\tilde{\alpha}_{13}-\tilde{\alpha}_{15})\hat{\Delta}_{[\mu\nu]\alpha}+\tilde{\alpha}_{1,11} \bar{\Delta}_{[\mu\nu]\alpha}+\tilde{\alpha}_{1,14}\mathring{\Delta}_{[\mu\nu]\alpha}\Big]\nonumber \\
		+\kappa \Big[ \tilde{\alpha}_{18}\breve{\Delta}_{\mu\nu\alpha}+ \tilde{\alpha}_{1,12}\bar{\Delta}_{\nu\alpha\mu}+\tilde{\alpha}_{1,13}\mathring{\Delta}_{\alpha\mu\nu} \Big]
	\end{gather}
	and
		\begin{gather}
	Q_{\nu\alpha\mu}=2 \kappa \Big[ (\tilde{\alpha}_{11}+\tilde{\alpha}_{16})\hat{\Delta}_{(\alpha\mu)\nu}+\tilde{\alpha}_{1,10} \bar{\Delta}_{(\alpha\mu)\nu}+\tilde{\alpha}_{1,13}\mathring{\Delta}_{(\alpha\mu)\nu}\Big] \nonumber \\
	+2 \kappa \Big[ (\tilde{\alpha}_{12}+\tilde{\alpha}_{15})\hat{\Delta}_{\nu(\alpha\mu)}+\tilde{\alpha}_{19} \breve{\Delta}_{\nu(\alpha\mu)}+\tilde{\alpha}_{1,12}\bar{\Delta}_{\nu(\alpha\mu)}\Big]\nonumber \\
	+2 \kappa \Big[ (\tilde{\alpha}_{13}+\tilde{\alpha}_{14})\hat{\Delta}_{(\mu|\nu|\alpha)}+\tilde{\alpha}_{18} \breve{\Delta}_{(\mu|\nu|\alpha)}+\tilde{\alpha}_{1,14}\mathring{\Delta}_{(\mu|\nu|\alpha)}\Big]
	\end{gather}
	respectively. In the above we have also exploited the anti-symmetry properties of $\breve{\Delta}_{\mu\nu\alpha}$, $\bar{\Delta}_{\mu\nu\alpha}$ and $\mathring{\Delta}_{\mu\nu\alpha}$ which follow immediately by their very definitions. Let us note that in the equation for torsion all elements $\tilde{\alpha}_{ij}$ of the inverse matrix appear, while in the expression for non-metricity the elements $\tilde{\alpha}_{17}$, $\tilde{\alpha}_{1,11}$ and $\tilde{\alpha}_{1,15}$ are absent. Again, the above expressions are exact and hold true for any kind of  hypermomentum. Of course, much more intuition will be gained once we specify the material content giving us the explicit form of the hypermomentum tensor.

	\section{Post Riemannian Expansion of the Metric Field Equations}

	Let us now elaborate on the metric field equations. Firstly, we take the trace of ($\ref{metricf}$) to arrive at
	\beq
	R+\mathcal{L}^{(2)}_{even}+\mathcal{L}^{(2)}_{odd}+\frac{1}{\sqrt{-g}}\partial_{\lambda}\Big( \sqrt{-g} \xi^{\lambda} \Big)=- \kappa T
	\eeq 
	where we have set
	\beq
	\xi^{\alpha}:=(c_{1}+n c_{2}+c_{3})S^{\alpha}+(2 a_{1}+2 n a_{3}+a_{5})Q^{\alpha}+(2 a_{2} +2 a_{4}+n a_{5})q^{\alpha}+(c_{6}+4 c_{4}+c_{5})t^{\alpha}
	\eeq
	In addition, using the post-Riemannian expansion of the Ricci scalar, the above trace equation becomes
	\begin{gather}
	\tilde{R}
	+	\Big( a_{1}+\frac{1}{4}\Big)  Q_{\alpha\mu\nu}Q^{\alpha\mu\nu} +
	\Big( a_{2}-\frac{1}{2}\Big)Q_{\alpha\mu\nu}Q^{\mu\nu\alpha} +
	\Big( a_{3}-\frac{1}{4}\Big)Q_{\mu}Q^{\mu}+
	a_{4}q_{\mu}q^{\mu}+
	\Big( a_{5}+\frac{1}{2}\Big)Q_{\mu}q^{\mu} \nonumber \\
	+(b_{1}+1)S_{\alpha\mu\nu}S^{\alpha\mu\nu} +
	(b_{2}-2)S_{\alpha\mu\nu}S^{\mu\nu\alpha} +
	(b_{3}-4)S_{\mu}S^{\mu}
	+	(c_{1}+2)Q_{\alpha\mu\nu}S^{\alpha\mu\nu}+
	(c_{2}-2)Q_{\mu}S^{\mu} +
	(c_{3}+2)q_{\mu}S^{\mu} \nonumber \\
	+a_{6}\varepsilon^{\alpha\beta\gamma\delta} Q_{\alpha\beta\mu}Q_{\gamma\delta}^{\;\;\;\;\mu} +b_{6}\varepsilon^{\alpha\beta\gamma\delta} S_{\alpha\beta\mu}S_{\gamma\delta}^{\;\;\;\;\mu}
	+c_{6}\varepsilon^{\alpha\beta\gamma\delta}Q_{\alpha\beta\mu}S_{\gamma\delta}^{\;\;\;\;\mu}+b_{5}S_{\mu}t^{\mu}+c_{4}Q_{\mu}t^{\mu}+c_{5}q^{\mu}t_{\mu} \nonumber \\
	+\frac{1}{\sqrt{-g}}\partial_{\mu}\Big[ \sqrt{-g}\Big( (2a_{2}+ 2 a_{4}+n a_{5}+1)q^{\mu}+(2 a_{1}+2 n a_{3}+a_{5}-1)Q^{\mu}\nonumber \\
	+(c_{1}+n c_{2}+c_{3}-4)S^{\mu} +(c_{6}+4 c_{4}+c_{5})t^{\mu}\Big) \Big] =
	-\kappa T \label{effEin}
	\end{gather}
	Notice that there exists non-trivial parameter space for which all the parity even quadratic terms disappear, this is of course the case of the teleparallel equivalents of GR \cite{jimenez2019general}. On the other hand, there is no parameter choice (except the trivial of vanishing coefficients) for which the parity odd terms cancel. Now, using a post-Riemannian expansion for every term in ($\ref{metricf}$), after some lengthy calculations (see appendix for details) we finally arrive at
	\begin{gather}
	\Big( \tilde{R}_{\mu\nu}-\frac{\tilde{R}}{2}g_{\mu\nu}\Big)=\kappa T_{\mu\nu} -\tilde{\nabla}_{\alpha}N^{\alpha}_{\;\;\;(\mu\nu)}+\tilde{\nabla}_{(\nu}N^{(2)}_{\mu)}-N^{(2)}_{\lambda}N^{\lambda}_{\;\;\;(\mu\nu)}+N^{\alpha}_{\;\;\;\lambda(\nu}N^{\lambda}_{\;\;\;\mu)\alpha}\nonumber \\
	+\frac{1}{2}g_{\mu\nu}\Big[ \tilde{\nabla}_{\alpha}( N^{(3)\alpha}-N^{(2)\alpha})+N^{(3)}_{\alpha}N^{(2)\alpha}-N_{\alpha\beta\gamma}N^{\beta\gamma\alpha} \Big] \nonumber \\
	+\frac{g_{\mu\nu}}{2} \Big[ \Big( \frac{b_{1}}{2}+ 2 a_{1}-c_{1}\Big) N_{\alpha\beta\gamma}N^{\alpha\beta\gamma}+	\Big( -\frac{b_{1}}{2}+  a_{2}+c_{1}\Big)N_{\alpha\beta\gamma}N^{\alpha\gamma\beta} \nonumber \\
	+	\Big( \frac{b_{2}}{2}+ 2 a_{2}+c_{1}\Big)N_{\alpha\beta\gamma}N^{\beta\gamma\alpha}+	\Big( -\frac{b_{2}}{2}+ 2 a_{1}+a_{2}-c_{1}\Big)N_{\kappa\lambda\alpha}N^{\lambda\kappa\alpha} \nonumber \\
	+\Big( \frac{b_{3}}{4}+4 a_{3}-c_{2} \Big)N^{(1)}_{\alpha}N^{(1)}_{\beta}g^{\alpha\beta}+\Big( \frac{b_{3}}{4}+a_{4}+\frac{c_{3}}{4} \Big)N^{(2)}_{\alpha}N^{(2)}_{\beta}g^{\alpha\beta}+a _{4}N^{(3)}_{\alpha}N^{(3)}_{\beta}g^{\alpha\beta} \nonumber \\
	+\Big( 2 a_{5}-\frac{c_{3}}{2}+c_{2}-\frac{b_{3}}{2}\Big) N^{(1)}_{\alpha}N^{(2)}_{\beta}g^{\alpha\beta}	+\Big( 2 a_{4}+\frac{c_{3}}{2}\Big)N^{(2)}_{\alpha}N^{(3)}_{\beta}g^{\alpha\beta}+\Big( 2 a_{5}-\frac{c_{3}}{2}\Big) N^{(1)}_{\alpha}N^{(3)}_{\beta}g^{\alpha\beta} \Big] \nonumber \\
	- \Big( 4 a_{1}-\frac{c_{1}}{2}\Big)\frac{1}{\sqrt{-g}}\hat{\nabla}_{\beta}\Big(\sqrt{-g}g^{\alpha\beta} N_{(\mu\nu)\alpha}\Big)-\Big( 2 a_{2}+\frac{c_{1}}{2}\Big)\frac{1}{\sqrt{-g}}\hat{\nabla}_{\beta}\Big(\sqrt{-g}g^{\alpha\beta} N_{(\mu|\alpha|\nu)}\Big)-2 a_{2}\frac{1}{\sqrt{-g}}\hat{\nabla}_{\beta}\Big(\sqrt{-g}g^{\alpha\beta}N_{\alpha(\mu\nu)}\Big) \nonumber \\
	-\frac{1}{\sqrt{-g}}\hat{\nabla}_{\alpha}\left[ \Big( 4 a_{3}-\frac{c_{2}}{2}\Big)N^{\alpha(1)}g_{\mu\nu}+\Big(  a_{5}+\frac{c_{2}}{2}\Big)N^{\alpha(2)}g_{\mu\nu}+a_{5}N^{\alpha(3)}g_{\mu\nu}\right]  \nonumber \\
	-\frac{1}{\sqrt{-g}}	\hat{\nabla}_{(\mu}  \left[ \Big( 2 a_{5}-\frac{c_{3}}{2}\Big)N_{\nu)}^{(1)}+\Big(  2 a_{4}+\frac{c_{3}}{2}\Big)N_{\nu)}^{(2)}+2 a_{4}N_{\nu)}^{(3)}\right]  \nonumber \\
	-\Big( 2 a_{1}-\frac{c_{1}}{2}\Big) N^{\alpha\beta}_{\;\;\;\;\mu}N_{\alpha\beta\nu}-\Big( 2 a_{1}-\frac{c_{1}}{2}\Big) N^{\alpha\beta}_{\;\;\;\;\mu}N_{\beta\alpha\nu}-\frac{c_{1}}{2}N^{\alpha\beta}_{\;\;\;\;(\mu}(N_{\beta|\nu)\alpha}+N_{\alpha|\nu)\beta} ) \nonumber \\
	+(N^{\alpha \;\;\beta}_{\;\;(\nu}+N_{(\nu}^{\;\; \alpha\beta})\left[ 2 a_{1}N_{\alpha|\mu)\beta}+ a_{2}N_{\beta|\mu)\alpha}+\Big( 2 a_{1}-\frac{c_{1}}{2}\Big)N_{\mu)\alpha\beta}+\Big( a_{2}+\frac{c_{1}}{2}\Big) N_{\mu)\beta\alpha} \right] \nonumber \\
	-\frac{1}{4}(N_{\beta(\nu|\alpha}-N_{\beta\alpha(\nu})\Big( 2 b_{1}(N^{\beta \;\;\alpha}_{\;\;\mu)}-b_{2}N^{\alpha \;\;\beta}_{\;\;\mu)}-2 b_{1}N^{\beta\alpha}_{\;\;\;\mu)}+ b_{2}N^{\alpha\beta}_{\;\;\;\mu)}\Big)+b_{1}N_{\mu\alpha\beta}N_{\nu}^{\;\;[\alpha\beta]} \nonumber \\
	+(N_{\mu\nu\alpha}+N_{\nu\mu\alpha})\left[ \Big( 4 a_{3}-\frac{c_{2}}{2}\Big) N^{(1)\alpha}+\Big( a_{5}+\frac{c_{2}}{2} \Big) N^{(2)\alpha}+ a_{5}N^{(3)\alpha} \right] \nonumber \\
	-\Big( 4 a_{3}-\frac{c_{3}}{2}\Big) N_{\mu}^{(1)}N_{\nu}^{(1)}-\Big( \frac{b_{3}}{4}-a_{4}\Big)  N_{\mu}^{(2)}N_{\nu}^{(2)}+a_{4} N_{\mu}^{(3)}N_{\nu}^{(3)}
	-\Big( c_{3}-\frac{b_{3}}{2}\Big)  N_{(\mu}^{(1)}N_{\nu)}^{(2)}+ 2 a_{4}  N_{(\mu}^{(2)}N_{\nu)}^{(3)} \nonumber \\
	-b_{6} \varepsilon^{\alpha\beta\gamma\delta} N_{\alpha\beta\nu}N_{\gamma\delta\mu}+ 2 a_{6} 	\varepsilon^{\alpha\beta\gamma\delta} N_{(\alpha\mu)\beta}(N_{\gamma\nu\delta}+N_{\nu\gamma\delta}) \nonumber \\
	- g_{\beta(\mu}g_{\nu)\lambda}\frac{1}{\sqrt{-g}}(\nabla_{\alpha}- 2 S_{\alpha})\Big(\sqrt{-g}  (c_{6}- 2 a_{6})\varepsilon^{\alpha\beta\gamma\delta} N^{\lambda}_{\;\;\gamma\delta}- 2 a_{6}\sqrt{-g} \varepsilon^{\alpha\beta\gamma\delta} N_{\gamma \;\;\;\;\delta}^{\;\;\lambda} \Big)
	\nonumber \\
	-\varepsilon^{\lambda\alpha\beta}_{\;\;\;\;\;\;(\nu}N_{\mu)\alpha\beta}\left[ \Big( 2 c_{4}-\frac{b_{5}}{2}\Big) N^{(1)}_{\lambda}+\Big(  c_{5}+\frac{b_{5}}{2}\Big) N^{(2)}_{\lambda}+ c_{5}N^{(3)}_{\lambda}\right] 
	-c_{4}g_{\mu\nu}\Big(\partial_{\lambda}(\sqrt{-g}M^{\lambda})\Big)
	\nonumber \\
	-c_{5}g_{\alpha(\mu}g_{\nu)\beta}\frac{1}{\sqrt{-g}}\nabla_{\lambda}(\sqrt{-g}M^{\alpha}g^{\lambda\beta})+ c_{5} \Big(N^{(2)}_{(\mu}-N^{(1)}_{(\mu}\Big)M_{\nu)}+\frac{c_{5}}{2}	\Big( N_{\nu\alpha\mu}+N_{\alpha\nu\mu}+N_{\mu\alpha\nu}+N_{\alpha\mu\nu}\Big) M^{\alpha} 
	 \label{GReff}
	\end{gather}
	and recall that
	\begin{gather}
N_{\alpha\mu\nu}=\kappa\tilde{\alpha}_{11}\hat{\Delta}_{\alpha\mu\nu}+\kappa\tilde{\alpha}_{12}\hat{\Delta}_{\nu\alpha\mu}+\kappa\tilde{\alpha}_{13} \hat{\Delta}_{\mu\nu\alpha}+\kappa\tilde{\alpha}_{14} \hat{\Delta}_{\alpha\nu\mu}+\kappa\tilde{\alpha}_{15} \hat{\Delta}_{\nu\mu\alpha} +\kappa\tilde{\alpha}_{16} \hat{\Delta}_{\mu\alpha\nu}+\kappa\tilde{\alpha}_{17}\breve{\Delta}_{\alpha\mu\nu}
+\kappa\tilde{\alpha}_{18} \breve{\Delta}_{\mu\nu\alpha} \nonumber \\ + \kappa\tilde{\alpha}_{19}\breve{\Delta}_{\nu\alpha\mu} 
+\kappa\tilde{\alpha}_{1,10}\bar{\Delta}_{\alpha\mu\nu}+\kappa\tilde{\alpha}_{1,11}\bar{\Delta}_{\mu\nu\alpha}+\kappa\tilde{\alpha}_{1,12}\bar{\Delta}_{\nu\alpha\mu}+\kappa\tilde{\alpha}_{1,13}\mathring{\Delta}_{\alpha\mu\nu}+\kappa\tilde{\alpha}_{1,14}\mathring{\Delta}_{\mu\nu\alpha}+\kappa\tilde{\alpha}_{1,15}\mathring{\Delta}_{\nu\alpha\mu} \label{MasterNB2}
\end{gather}
	where the various 
	$\Delta_{\alpha\mu\nu}'s$ are given by the expressions $(\ref{B1})$-$(\ref{B4})$ by setting $B_{\alpha\mu\nu}=\kappa \Delta_{\alpha\mu\nu}$. The field equations $(\ref{GReff})$ represent a modified version of GR with additional matter couplings that are associated to the microscopic characteristics of matter. It is obvious then, by direct substitution of the latter form of the distortion into $(\ref{GReff})$ that apart from the usual energy-momentum tensor quadratic and derivative terms of the hypermomentum serve also as sources producing gravitational fields. Our result above serves as a generalization of the earlier findings of \cite{Iosifidis:2021tvx} where now the additional $6$ parity odd quadratic scalars have been added into the action along with the $11$  parity even ones.  Our study now covers the whole spectrum of quadratic Metric-Affine Gravity Theories, that is we have found the exact form of the connection and consequently that of torsion and non-metricity for the most general $17$ parameter quadratic MAG action.
	
	 Of course, given the complexity of the right-hand side of the above field equations, no concrete results can be inferred. More information can be extracted once a specific matter configuration is adopted giving rise to a certain form of the hypermomentum tensor. However, one thing one can be sure about is that the effective energy-momentum tensor  as given by the right-hand side of ($\ref{GReff}$), that is, the sum of usual energy-momentum tensor $+$ hypermomentum contributions is covariantly conserved with respect to the Levi-Civita connection. Indeed, since the Riemannian contracted Bianchi  identities ensure that the left hand side of the above field equations is covariantly conserved with respect  to the Levi-Civita connection so must be for the right-hand side as well. 
	
	Now, it is interesting to study what happens when the matter fields do not couple to the connection, the so-called Palatini case. In this instance $\Delta_{\alpha\mu\nu}\equiv 0$ and from corollary $1$ we also conclude the following
	\begin{corollary}
		In the Palatini version of ($\ref{action}$), that is when the matter sector of the action is independent of the connection (i.e. $\Delta_{\alpha\mu\nu}\equiv 0$) and given that both $det(A)\neq 0$ and $det(\Gamma)\neq 0$ the connection solution is always the Levi-Civita. In other words, torsion and non-metricity vanish identically and the $17$ parameter quadratic Theory is equivalent to GR. 
	\end{corollary}
A special case of the latter is of course the vacuum Theory and therefore we have also the following result holding true.
	\begin{corollary}
		In vacuum  and given that both matrices $A$  and $\Gamma$ are non-singular,  the $17$ parameter quadratic Theory always reduces to GR. 
	\end{corollary}
It is rather remarkable that despite the complexity of the full quadratic Theory, in vacuum the latter is equivalent to GR. Important deviations occur only when matter with intrinsic structure is present, that is matter with non-vanishing hypermomentum.

Now, let us stress out that in applying the results of $Theorem-1$ in order to solve for the distortion, nowhere did we assume that the coefficients $a_{i}$, $b_{j}$ and $c_{k}$ are constants. Consequently,, the main result for the affine connection will continue to hold true even if we  promote the aforementioned constants to be spacetime functions. In particular, the following generalized Theorem holds true.

\begin{theo}
	Let $\Phi$ be a collection of matter fields (collectively denoted) and $d\Phi$ their exterior derivatives. Furthermore, let $a_{i}(\Phi, d \Phi)=a_{i}(\Phi, \partial \Phi)$, $b_{i}(\Phi, \partial \Phi)$ and $c_{i}(\Phi, \partial \Phi)$ $i=1,2,...,6$ be arbitrary functions continuous and differentiable in their arguments\footnote{These functions can just as well depend also on the metric tensor and its derivatives, but not on the connection.}. Then, for the general class of Theories given by the action
	\begin{gather}
	S[g, \Gamma, \Phi]
	=\frac{1}{2 \kappa}\int d^{n}x \sqrt{-g} \Big[  R+ 
	b_{1}(\Phi, \partial \Phi)S_{\alpha\mu\nu}S^{\alpha\mu\nu} +
	b_{2}(\Phi, \partial \Phi)S_{\alpha\mu\nu}S^{\mu\nu\alpha} +
	b_{3}(\Phi, \partial \Phi)S_{\mu}S^{\mu} \nonumber \\
	a_{1}(\Phi, \partial \Phi)Q_{\alpha\mu\nu}Q^{\alpha\mu\nu} +
	a_{2}(\Phi, \partial \Phi)Q_{\alpha\mu\nu}Q^{\mu\nu\alpha} +
	a_{3}(\Phi, \partial \Phi)Q_{\mu}Q^{\mu}+
	a_{4}(\Phi, \partial \Phi)q_{\mu}q^{\mu}\nonumber \\
+
a_{5}(\Phi, \partial \Phi)Q_{\mu}q^{\mu} 	+c_{1}(\Phi, \partial \Phi)Q_{\alpha\mu\nu}S^{\alpha\mu\nu}+
	c_{2}(\Phi, \partial \Phi)Q_{\mu}S^{\mu} +
	c_{3}(\Phi, \partial \Phi)q_{\mu}S^{\mu}  \nonumber \\
		+a_{6}(\Phi, \partial \Phi)\varepsilon^{\alpha\beta\gamma\delta}Q_{\alpha\beta\mu}Q_{\gamma\delta}^{\;\;\;\;\mu}+b_{5}(\Phi, \partial \Phi)S_{\mu}t^{\mu}+b_{6}(\Phi, \partial \Phi)\varepsilon^{\alpha\beta\gamma\delta}S_{\alpha\beta\mu}S_{\gamma\delta}^{\;\;\;\;\mu} \nonumber \\
	c_{4}(\Phi, \partial \Phi)Q_{\mu}t^{\mu}+c_{5}(\Phi, \partial \Phi)q^{\mu}t_{\mu}+c_{6}\varepsilon^{\alpha\beta\gamma\delta}Q_{\alpha\beta\mu}S_{\gamma\delta}^{\;\;\;\;\mu}
	\Big] +S_{M}[g, \Gamma, \Phi]  \label{genquad2}
	\end{gather}	
	the exact solution for the distortion tensor reads
		\begin{gather}
	N_{\alpha\mu\nu}=\tilde{\alpha}_{11}\hat{B}_{\alpha\mu\nu}+\tilde{\alpha}_{12}\hat{B}_{\nu\alpha\mu}+\tilde{\alpha}_{13} \hat{B}_{\mu\nu\alpha}+\tilde{\alpha}_{14} \hat{B}_{\alpha\nu\mu}+\tilde{\alpha}_{15} \hat{B}_{\nu\mu\alpha} +\tilde{\alpha}_{16} \hat{B}_{\mu\alpha\nu}+\tilde{\alpha}_{17}\breve{B}_{\alpha\mu\nu}+\tilde{\alpha}_{18} \breve{B}_{\mu\nu\alpha} + \tilde{\alpha}_{19}\breve{B}_{\nu\alpha\mu} \nonumber \\
	+\tilde{\alpha}_{1,10}\bar{B}_{\alpha\mu\nu}+\tilde{\alpha}_{1,11}\bar{B}_{\mu\nu\alpha}+\tilde{\alpha}_{1,12}\bar{B}_{\nu\alpha\mu}+\tilde{\alpha}_{1,13}\mathring{B}_{\alpha\mu\nu}+\tilde{\alpha}_{1,14}\mathring{B}_{\mu\nu\alpha}+\tilde{\alpha}_{1,15}\mathring{B}_{\nu\alpha\mu} 
	\end{gather}
	
and consequently	the affine connection is given by
	\begin{gather}
\Gamma^{\lambda}_{\;\;\;\mu\nu}	=\tilde{\Gamma}^{\lambda}_{\;\;\;\mu\nu}+\tilde{\alpha}_{11}\hat{B}_{\alpha\mu\nu}+\tilde{\alpha}_{12}\hat{B}_{\nu\alpha\mu}+\tilde{\alpha}_{13} \hat{B}_{\mu\nu\alpha}+\tilde{\alpha}_{14} \hat{B}_{\alpha\nu\mu}+\tilde{\alpha}_{15} \hat{B}_{\nu\mu\alpha} +\tilde{\alpha}_{16} \hat{B}_{\mu\alpha\nu}+\tilde{\alpha}_{17}\breve{B}_{\alpha\mu\nu}+\tilde{\alpha}_{18} \breve{B}_{\mu\nu\alpha} + \tilde{\alpha}_{19}\breve{B}_{\nu\alpha\mu} \nonumber \\
	+\tilde{\alpha}_{1,10}\bar{B}_{\alpha\mu\nu}+\tilde{\alpha}_{1,11}\bar{B}_{\mu\nu\alpha}+\tilde{\alpha}_{1,12}\bar{B}_{\nu\alpha\mu}+\tilde{\alpha}_{1,13}\mathring{B}_{\alpha\mu\nu}+\tilde{\alpha}_{1,14}\mathring{B}_{\mu\nu\alpha}+\tilde{\alpha}_{1,15}\mathring{B}_{\nu\alpha\mu} \label{co}
	\end{gather}
	where now $\tilde{\alpha}_{ij}=\tilde{\alpha}_{ij}(\Phi, \partial \Phi)$.

\end{theo}

\textit{Proof.}	Following an identical procedure with that of section $V$ and by simply promoting the $a_{i}'s, b_{j}'s$ and $c_{j}'s$ to be functions of $\Phi$ and $d\Phi$ we arrive at the stated result. The last action covers a wide range of Metric-Affine Gravity Theories and includes as special cases non-minimally coupled scalar-tensor and vector-tensor Theories.

	\section{Conclusions}

	We have studied the most general MAG Theory that is quadratic in torsion and non-metricity (and their mixings) and linear in the curvature. Without restricting ourselves to the parity even case, we included all the parity odd terms as well. For this $17$ parameter Theory, we first derived the field equations by varying independently with respect to the metric as well as the connection. Focusing first on the connection field equations, we performed a simple trick to recast the set of these equations solely in terms of the distortion instead of combinations of torsion and non-metricity. Then, using a recent result on solving linear $30$ parameter tensor equations (see ref. \cite{Iosifidis:2021tvx}) we were able to find the exact and unique form of the distortion and subsequently of the affine connection in terms of the hypermomentum and its various contractions. The exact forms of torsion and non-metricity can then be derived readily. Continuing, we turned our attention to the metric field equations and by using a post Riemannian expansion for all relevant pieces we recast the latter into an effective GR with modified source terms. These extended sources, apart from the usual energy-momentum contribution, are comprised by quadratic combinations of the hypermomentum tensor, its dual constructions as well as linear derivatives of both the former and the latter tensors and their contractions. Their physical role is tightly connected to the microstructure of matter. The exact effect is given by equation $(\ref{GReff})$ but given its complexity no concrete conclusions can be inferred without knowing more about the form of the sources. Equation ($\ref{GReff}$) , as it stands,  provides a generalization of Einstein-Cartan Theory where on top of spacetime torsion sourced by the spin part of hypermomentum, the dilation and shear parts of the latter excite also non-metricity producing, therefore, a fully non-Riemannian arena for Gravity.
	
	Now, even though we are able to obtain the full solution in the presence of the sources, of equal importance is to know what the vacuum case corresponds to. In other words, the question:  "What the full quadratic Theory (as given by action $(\ref{S})$) corresponds to in vacuum? " was not known till know. In the current research we were able to answer to this question as well and as it turns out the vacuum Theory, despite all appearances, is equivalent to GR. As a result there are no new (and possible harmful) degrees of freedom and the new interactions are tied to matter. In addition, not only the vacuum Theory but also the less restricted Palatini case where the matter does not depend on the connection (i.e. identically vanishing hypermomentum) but does depend on the metric, yields again GR. It also becomes apparent that in order to fully appreciate, and be able to deeply study the role of Metric-Affine Theories of Gravity, the inclusion of hypermomentum is all essential and indispensable. In other words, the connection-matter couplings are of prominent importance and have to be taken seriously \cite{Hehl:1977gn}.  After all it is exactly the role of the latter couplings that distinguish MAG from the rest of modified gravity Theories, which also provides the astonishing relation between the microstructure of matter and the excitations of spacetime torsion and non-metricity. The intrinsic structure of matter is fully described by the hypermomentum tensor.
	
	In addition, as we explicitly showed, our result for the connection solution continues to hold true even for more general couplings for the quadratic invariants that are not necessary constant. In particular, we have also found the exact connection solution, as given by equation ($\ref{co}$), for the fairly  wide class of Theories ($\ref{genquad2}$). 
	Subsequently, the next logical step to apply the results we obtained here would be to consider specific forms for the hypermomentum and find the exact form of these extra source terms. Then, the  forms of torsion and non-metricity can be obtained and also the new terms on the right-hand side of $(\ref{GReff})$ could be interpreted. One could then perhaps consider the manifield construction of \cite{Cant:1985yg}. Additionally, it would also be very interesting to study the $17$ parameter quadratic Theory in the presence of an isotropic hyperfluid \cite{iosifidis2021perfect} or even the homogeneous Cosmology of the quadratic Theory sourced by the Perfect Cosmological Hyperfluid \cite{Iosifidis:2020gth}. Some of these aspects are under investigation now.

		\section{Acknowledgments}	This research is co-financed by Greece and the European Union (European Social Fund- ESF) through the
	Operational Programme 'Human Resources Development, Education and Lifelong Learning' in the context
	of the project “Reinforcement of Postdoctoral Researchers - 2
	nd Cycle” (MIS-5033021), implemented by the
	State Scholarships Foundation (IKY).

	\appendix

	\section{Variations}
	Let us gather here the variations of the parity odd terms with respect to the metric as well as the affine connection. Recall that the parity odd quadratic Lagrangian reads
	\begin{gather}
	\mathcal{L}_{odd}= a_{6}\varepsilon^{\alpha\beta\gamma\delta} Q_{\alpha\beta\mu}Q_{\gamma\delta}^{\;\;\;\;\mu} +b_{6}\varepsilon^{\alpha\beta\gamma\delta} S_{\alpha\beta\mu}S_{\gamma\delta}^{\;\;\;\;\mu}
	+c_{6}\varepsilon^{\alpha\beta\gamma\delta}Q_{\alpha\beta\mu}S_{\gamma\delta}^{\;\;\;\;\mu}+b_{5}S_{\mu}t^{\mu}+c_{4}Q_{\mu}t^{\mu}+c_{5}q^{\mu}t_{\mu}
	\end{gather}
	We then find the following.

	\subsection{g-Variations}
	
	\beq
	\delta_{g}\Big( \sqrt{-g} \varepsilon^{\alpha\beta\gamma\delta} Q_{\alpha\beta\mu}Q_{\gamma\delta}^{\;\;\;\;\mu} \Big)=-2 g_{\beta(\mu}g_{\nu)\lambda}(\nabla_{\alpha}- 2 S_{\alpha})\Big(\sqrt{-g}\varepsilon^{\alpha\beta\gamma\delta} Q_{\gamma\delta}^{\;\;\;\;\lambda}\Big) (\delta g^{\mu\nu})+\sqrt{-g}\varepsilon^{\alpha\beta\gamma\delta}Q_{\alpha\beta\mu}Q_{\gamma\delta\nu}(\delta g^{\mu\nu})
	\eeq
	\beq
	\delta_{g}\Big( \sqrt{-g} \varepsilon^{\alpha\beta\gamma\delta} S_{\alpha\beta\mu}S_{\gamma\delta}^{\;\;\;\;\mu} \Big)=-\sqrt{-g}  \varepsilon^{\alpha\beta\gamma\delta} S_{\alpha\beta\nu}S_{\gamma\delta\mu}(\delta g^{\mu\nu})
	\eeq
	\beq
	\delta_{g}\Big( \sqrt{-g} \varepsilon^{\alpha\beta\gamma\delta} Q_{\alpha\beta\mu}S_{\gamma\delta}^{\;\;\;\;\mu} \Big)=- g_{\beta(\mu}g_{\nu)\lambda}(\nabla_{\alpha}- 2 S_{\alpha})\Big(\sqrt{-g}\varepsilon^{\alpha\beta\gamma\delta} S_{\gamma\delta}^{\;\;\;\;\lambda}\Big) (\delta g^{\mu\nu})
	\eeq
	\beq
	\delta_{g}(\sqrt{-g}t^{\mu}S_{\mu})=-\sqrt{-g}S_{\alpha\beta\mu}\varepsilon^{\lambda\alpha\beta}_{\;\;\;\;\;\;\nu}S_{\lambda}(\delta g^{\mu\nu})
	\eeq
	
	\beq
	\delta_{g}(\sqrt{-g}t^{\lambda}Q_{\lambda})=-\sqrt{-g}S_{\alpha\beta\mu}\varepsilon^{\lambda\alpha\beta}_{\;\;\;\;\;\;\nu}Q_{\lambda}(\delta g^{\mu\nu})-g_{\mu\nu}\Big(\partial_{\lambda}(\sqrt{-g}t^{\lambda})\Big)(\delta g^{\mu\nu})+t.d.
	\eeq
	\begin{gather}
	\delta_{g}(\sqrt{-g}t^{\lambda}q_{\lambda})=\Big[ -\sqrt{-g}S_{\alpha\beta\mu}\varepsilon^{\lambda\alpha\beta}_{\;\;\;\;\;\;\nu}q_{\lambda}-g_{\mu\alpha}g_{\nu\beta}\nabla_{\lambda}(\sqrt{-g}t^{\alpha}g^{\lambda\beta})+ \sqrt{-g}2 S_{\mu}t_{\nu}+\sqrt{-g}t^{\alpha}Q_{\mu\nu\alpha} \Big] (\delta g^{\mu\nu})
	\end{gather}
	So up to total derivatives we find
	\begin{gather}
	\delta_{g}\Big(\sqrt{-g}\mathcal{L}^{(2)}_{odd}\Big)=\Big[-b_{6}\sqrt{-g}  \varepsilon^{\alpha\beta\gamma\delta} S_{\alpha\beta\nu}S_{\gamma\delta\mu}-2 g_{\beta(\mu}g_{\nu)\lambda}(\nabla_{\alpha}- 2 S_{\alpha})\Big(\sqrt{-g} a_{6} \varepsilon^{\alpha\beta\gamma\delta} Q_{\gamma\delta}^{\;\;\;\;\lambda}\Big)
	\nonumber \\
	+\sqrt{-g} a_{6}\varepsilon^{\alpha\beta\gamma\delta}Q_{\alpha\beta\mu}Q_{\gamma\delta\nu}- g_{\beta(\mu}g_{\nu)\lambda}(\nabla_{\alpha}- 2 S_{\alpha})\Big(\sqrt{-g}c_{6}\varepsilon^{\alpha\beta\gamma\delta} S_{\gamma\delta}^{\;\;\;\;\lambda}\Big)
	\nonumber \\
	-\sqrt{-g}b_{5}S_{\alpha\beta\mu}\varepsilon^{\lambda\alpha\beta}_{\;\;\;\;\;\;\nu}S_{\lambda}
	-\sqrt{-g}c_{4}S_{\alpha\beta\mu}\varepsilon^{\lambda\alpha\beta}_{\;\;\;\;\;\;\nu}Q_{\lambda}-c_{4}g_{\mu\nu}\Big(\partial_{\lambda}(\sqrt{-g}t^{\lambda})\Big)
	\nonumber \\
	-\sqrt{-g}c_{5}S_{\alpha\beta\mu}\varepsilon^{\lambda\alpha\beta}_{\;\;\;\;\;\;\nu}q_{\lambda}-c_{5}g_{\mu\alpha}g_{\nu\beta}\nabla_{\lambda}(\sqrt{-g}t^{\alpha}g^{\lambda\beta})+ \sqrt{-g}c_{5}2 S_{\mu}t_{\nu}+\sqrt{-g}c_{5}t^{\alpha}Q_{\mu\nu\alpha} \Big] (\delta g^{\mu\nu})
	\end{gather}
	and after some rearranging we finally arrive at
	\begin{gather}
	\frac{1}{\sqrt{-g}}	\frac{\Big(\sqrt{-g}\mathcal{L}^{(2)}_{odd}\Big)}{\delta g^{\mu\nu}}=-b_{6} \varepsilon^{\alpha\beta\gamma\delta} S_{\alpha\beta\nu}S_{\gamma\delta\mu}+ a_{6}\varepsilon^{\alpha\beta\gamma\delta}Q_{\alpha\beta\mu}Q_{\gamma\delta\nu} \nonumber \\
	- g_{\beta(\mu}g_{\nu)\lambda}\frac{1}{\sqrt{-g}}(\nabla_{\alpha}- 2 S_{\alpha})\Big(\sqrt{-g}  \varepsilon^{\alpha\beta\gamma\delta} ( 2 a_{6} Q_{\gamma\delta}^{\;\;\;\;\lambda}+c_{6} S_{\gamma\delta}^{\;\;\;\;\lambda}) \Big)
	\nonumber \\
	-S_{\alpha\beta(\mu}\varepsilon^{\lambda\alpha\beta}_{\;\;\;\;\;\;\nu)}(b_{5} S_{\lambda}+c_{4}Q_{\lambda}+c_{5} q_{\lambda})
	-c_{4}g_{\mu\nu}\Big(\partial_{\lambda}(\sqrt{-g}t^{\lambda})\Big)
	\nonumber \\
	-c_{5}g_{\alpha(\mu}g_{\nu)\beta}\frac{1}{\sqrt{-g}}\nabla_{\lambda}(\sqrt{-g}t^{\alpha}g^{\lambda\beta})+ c_{5}2 S_{(\mu}t_{\nu)}+c_{5}t^{\alpha}Q_{(\mu\nu)\alpha} \label{gpart}
	\end{gather}
	
	\subsection{$\Gamma$-Variations}

		\beq
	\delta_{\Gamma}\Big( \varepsilon^{\alpha\beta\gamma\delta} Q_{\alpha\beta\mu}Q_{\gamma\delta}^{\;\;\;\;\mu} \Big)=-2\Big( \varepsilon^{\mu\nu\alpha\beta}Q_{\alpha\beta\lambda}+\varepsilon_{\lambda}^{\;\;\nu\alpha\beta}Q_{\alpha\beta}^{\;\;\;\;\mu}\Big)\delta \Gamma^{\lambda}_{\;\;\;\mu\nu}
	\eeq
	\beq
	\delta_{\Gamma}\Big( \varepsilon^{\alpha\beta\gamma\delta} S_{\alpha\beta\mu}S_{\gamma\delta}^{\;\;\;\;\mu} \Big)=2 \varepsilon^{\mu\nu\alpha\beta}S_{\alpha\beta\lambda}\delta\Gamma^{\lambda}_{\;\;\;\mu\nu}
	\eeq
	\beq
	\delta_{\Gamma}\Big(\varepsilon^{\alpha\beta\gamma\delta}Q_{\alpha\beta\mu}S_{\gamma\delta}^{\;\;\;\;\mu} \Big)=\Big( -\varepsilon^{\mu\nu\alpha\beta}S_{\alpha\beta\lambda}-\varepsilon_{\lambda}^{\;\;\nu\alpha\beta}S_{\alpha\beta}^{\;\;\;\;\mu}+\varepsilon^{\mu\nu\alpha\beta}Q_{\alpha\beta\lambda}\Big)\delta \Gamma^{\lambda}_{\;\;\;\mu\nu}
	\eeq
	Let
	\beq
	B:=(b_{5}S_{\mu}+c_{4}Q_{\mu}+c_{5}q_{\mu})t^{\mu}
	\eeq
	Then
	\beq
	\delta_{\Gamma}B=\Big[ \varepsilon^{\alpha\mu\nu}_{\;\;\;\;\;\;\lambda}(b_{5}S_{\alpha}+c_{4}Q_{\alpha}+c_{5}q_{\alpha})+b_{5}t^{[\mu}\delta^{\nu]}_{\lambda}+2 c_{4}t^{\nu}\delta^{\mu}_{\lambda}+c_{5}(g^{\mu\nu}t_{\lambda}+t^{\mu}\delta^{\nu}_{\lambda})\Big]\delta \Gamma^{\lambda}_{\;\;\;\mu\nu}
	\eeq
	and as a result 
	the	$\Gamma$ variation of the parity odd Lagrangian reads
	\begin{gather}
	\delta_{\Gamma}\mathcal{L}_{odd}=\delta \Gamma^{\lambda}_{\;\;\;\mu\nu}\Big[ (-2 a_{6}+c_{6})\varepsilon^{\mu\nu\alpha\beta}Q_{\alpha\beta\lambda}+(2 b_{6}-c_{6})\varepsilon^{\mu\nu\alpha\beta}S_{\alpha\beta\lambda}-2 a_{6}\varepsilon_{\lambda}^{\;\;\nu\alpha\beta}Q_{\alpha\beta}^{\;\;\;\;\mu}-c_{6}\varepsilon_{\lambda}^{\;\;\nu\alpha\beta}S_{\alpha\beta}^{\;\;\;\;\mu} \nonumber \\
	+\varepsilon^{\alpha\mu\nu}_{\;\;\;\;\;\;\lambda}(b_{5}S_{\alpha}+c_{4}Q_{\alpha}+c_{5}q_{\alpha})+\Big( \frac{b_{5}}{2}+c_{5}\Big) t^{\mu}\delta_{\lambda}^{\nu}+\Big( -\frac{b_{5}}{2}+2c_{4}\Big) t^{\nu}\delta_{\lambda}^{\mu}+c_{5}g^{\mu\nu}t_{\lambda} \Big]
	\end{gather}

	\subsection{More Variations}
	We compute
	\beq
	g^{\mu\nu}	\frac{\delta\Big(	\sqrt{-g}	\varepsilon^{\alpha\beta\gamma\delta}S_{\alpha\beta\mu}S_{\gamma\delta}^{\;\;\;\;\mu} \Big)}{\delta g^{\mu\nu}}=-\sqrt{-g}		\varepsilon^{\alpha\beta\gamma\delta}S_{\alpha\beta\mu}S_{\gamma\delta}^{\;\;\;\;\mu}
	\eeq
	\beq
	g^{\mu\nu} \frac{\delta	\Big( \sqrt{-g} \varepsilon^{\alpha\beta\gamma\delta} Q_{\alpha\beta\mu}S_{\gamma\delta}^{\;\;\;\;\mu} \Big)}{\delta g^{\mu\nu}}= -\sqrt{-g} \varepsilon^{\alpha\beta\gamma\delta} Q_{\alpha\beta\mu}S_{\gamma\delta}^{\;\;\;\;\mu} -\partial_{\lambda}(\sqrt{-g}t^{\lambda})
	\eeq
	\beq
	g^{\mu\nu}	\frac{\delta\Big(	\sqrt{-g}	\varepsilon^{\alpha\beta\gamma\delta}Q_{\alpha\beta\mu}Q_{\gamma\delta}^{\;\;\;\;\mu} \Big)}{\delta g^{\mu\nu}}=-\sqrt{-g}		\varepsilon^{\alpha\beta\gamma\delta}Q_{\alpha\beta\mu}Q_{\gamma\delta}^{\;\;\;\;\mu}
	\eeq
	\beq
	g^{\mu\nu}	\frac{\delta\Big( \sqrt{-g}t^{\alpha} S_{\alpha} \Big)}{\delta g^{\mu\nu}}=	-\sqrt{-g}t^{\alpha} S_{\alpha}
	\eeq
	\beq
	g^{\mu\nu}	\frac{\delta\Big( \sqrt{-g}t^{\alpha} q_{\alpha} \Big)}{\delta g^{\mu\nu}}=	-\sqrt{-g}t^{\alpha} q_{\alpha}-\partial_{\lambda}(\sqrt{-g}t^{\lambda})
	\eeq
	\beq
	g^{\mu\nu}	\frac{\delta\Big( \sqrt{-g}t^{\alpha} Q_{\alpha} \Big)}{\delta g^{\mu\nu}}=	-\sqrt{-g}t^{\alpha} Q_{\alpha}- 4 \partial_{\lambda}(\sqrt{-g}t^{\lambda})
	\eeq
	With these we easily find that
	\beq
	g^{\mu\nu}	\frac{\delta\Big(	\sqrt{-g} \mathcal{L}^{(2)}_{odd}\Big)}{\delta g^{\mu\nu}}=-\sqrt{-g} \mathcal{L}^{(2)}_{odd}-(c_{6}+4 c_{4}+c_{5})\partial_{\lambda}(\sqrt{-g}t^{\lambda})
	\eeq	
	Note that for the parameter choice $c_{6}+4 c_{4}+c_{5}=0$ the derivative term disappears.

	\section{Expansions}
	
	\subsection{Distortion vectors in terms of torsion and non-metricity vectors}
	Contracting $N_{\alpha\mu\nu}$ in all possible ways with the metric  we find
	\beq
	N_{(1)}^{\mu}=\frac{1}{2}Q^{\mu}\; \;, \;\; N_{(2)}^{\mu}=\frac{1}{2}Q^{\mu}+2 S^{\mu}\;\;, \;\; N_{(3)}^{\mu}=q^{\mu}-\frac{1}{2}Q^{\mu}-2 S^{\mu}
	\eeq
	which relate the distortion vectors in terms of those of torsion and non-metricity. Inverting them we have
	\beq
	Q^{\mu}=2 N_{(1)}^{\mu}\;\;, \;\; q^{\mu}=N_{(2)}^{\mu}+N_{(3)}^{\mu}\;\;, \;\; S^{\mu}=\frac{1}{2}\Big( N_{(2)}^{\mu}-N_{(1)}^{\mu}\Big)
	\eeq
	Then,  upon using ($\ref{QNSN}$) along with the above relations and the definition of the Palatini tensor
		\beq
	P_{\lambda}^{\;\;\;\mu\nu}=-\frac{\nabla_{\lambda}(\sqrt{-g}g^{\mu\nu})}{\sqrt{-g}}+\frac{\nabla_{\sigma}(\sqrt{-g}g^{\mu\sigma})\delta^{\nu}_{\lambda}}{\sqrt{-g}} \\
	+2(S_{\lambda}g^{\mu\nu}-S^{\mu}\delta_{\lambda}^{\nu}+g^{\mu\sigma}S_{\sigma\lambda}^{\;\;\;\;\nu})
	\eeq
	 we can express the latter solely in terms of the distortion as
	\beq
	P^{\alpha\mu\nu}=g^{\mu\nu}N_{(2)}^{\alpha}+g^{\nu\alpha}N_{(3)}^{\mu}-(N^{\mu\nu\alpha}+N^{\nu\alpha\mu})
	\eeq
	
	\subsection{Post-Riemannian expansions}
	The Ricci tensor is expanded as
		\begin{equation}
	R_{\nu\beta}=\tilde{R}_{\nu\beta}+ \tilde{\nabla}_{\mu}N^{\mu}_{\;\;\;\nu\beta}-\tilde{\nabla}_{\beta}N^{(2)}_{\nu}+N^{(2)}_{\lambda}N^{\lambda}_{\;\;\;\nu\beta}-N^{\mu}_{\;\;\;\rho\beta}N^{\rho}_{\;\;\;\nu\mu}
	\end{equation}
and the Ricci scalar as
	\begin{equation}
	R=\tilde{R}+ \tilde{\nabla}_{\mu}( N^{(3)\mu}-N^{(2)\mu})+ N^{(3)}_{\mu}N^{(2)\mu}-N_{\alpha\mu\nu}N^{\mu\nu\alpha} \label{Recomp}
	\end{equation}
		As a result for the Einstein-like combination we find
		\begin{gather}
		R_{(\mu\nu)}-\frac{R}{2}g_{\mu\nu}=	\Big( \tilde{R}_{(\mu\nu)}-\frac{\tilde{R}}{2}g_{\mu\nu}\Big)+\tilde{\nabla}_{\alpha}N^{\alpha}_{\;\;\;(\mu\nu)}-\tilde{\nabla}_{(\nu}N^{(2)}_{\mu)}+N^{(2)}_{\lambda}N^{\lambda}_{\;\;\;(\mu\nu)}-N^{\alpha}_{\;\;\;\lambda(\nu}N^{\lambda}_{\;\;\;\mu)\alpha}\nonumber \\
		-\frac{1}{2}g_{\mu\nu}\Big[ \tilde{\nabla}_{\alpha}( N^{(3)\alpha}-N^{(2)\alpha})+N^{(3)}_{\alpha}N^{(2)\alpha}-N_{\alpha\beta\gamma}N^{\beta\gamma\alpha} \Big]
		\end{gather}
		Continuing we calculate
		\begin{gather}
			b_{1}S_{\alpha\mu\nu}S^{\alpha\mu\nu} +
		b_{2}S_{\alpha\mu\nu}S^{\mu\nu\alpha} +
		a_{1}Q_{\alpha\mu\nu}Q^{\alpha\mu\nu} +
		a_{2}Q_{\alpha\mu\nu}Q^{\mu\nu\alpha} +
		+c_{1}Q_{\alpha\mu\nu}S^{\alpha\mu\nu}= \nonumber \\
		\Big( \frac{b_{1}}{2}+ 2 a_{1}-c_{1}\Big) N_{\alpha\mu\nu}N^{\alpha\mu\nu}+	\Big( -\frac{b_{1}}{2}+  a_{2}+c_{1}\Big)N_{\alpha\mu\nu}N^{\alpha\nu\mu} \nonumber \\
		+	\Big( \frac{b_{2}}{2}+ 2 a_{2}+c_{1}\Big)N_{\alpha\mu\nu}N^{\mu\nu\alpha}+	\Big( -\frac{b_{2}}{2}+ 2 a_{1}+a_{2}-c_{1}\Big)N_{\mu\nu\alpha}N^{\nu\mu\alpha}
		\end{gather}
		and also
			\begin{gather}
		b_{3}S_{\mu}S^{\mu} +
		a_{3}Q_{\mu}Q^{\mu}+
		a_{4}q_{\mu}q^{\mu}+
		a_{5}Q_{\mu}q^{\mu}+
		c_{2}Q_{\mu}S^{\mu} +
		c_{3}q_{\mu}S^{\mu}= \nonumber \\
		=\Big( \frac{b_{3}}{4}+4 a_{3}-c_{2} \Big)N^{(1)}_{\mu}N^{(1)}_{\nu}g^{\mu\nu}+\Big( \frac{b_{3}}{4}+a_{4}+\frac{c_{3}}{4} \Big)N^{(2)}_{\mu}N^{(2)}_{\nu}g^{\mu\nu}+a _{4}N^{(3)}_{\mu}N^{(3)}_{\nu}g^{\mu\nu} \nonumber \\
	+\Big( 2 a_{5}-\frac{c_{3}}{2}+c_{2}-\frac{b_{3}}{2}\Big) N^{(1)}_{\mu}N^{(2)}_{\nu}g^{\mu\nu}	+\Big( 2 a_{4}+\frac{c_{3}}{2}\Big)N^{(2)}_{\mu}N^{(3)}_{\nu}g^{\mu\nu}+\Big( 2 a_{5}-\frac{c_{3}}{2}\Big) N^{(1)}_{\mu}N^{(3)}_{\nu}g^{\mu\nu}
		\end{gather}
		such that
			\begin{gather}
	\mathcal{L}_{2}=\Big( \frac{b_{1}}{2}+ 2 a_{1}-c_{1}\Big) N_{\alpha\mu\nu}N^{\alpha\mu\nu}+	\Big( -\frac{b_{1}}{2}+  a_{2}+c_{1}\Big)N_{\alpha\mu\nu}N^{\alpha\nu\mu} \nonumber \\
	+	\Big( \frac{b_{2}}{2}+ 2 a_{2}+c_{1}\Big)N_{\alpha\mu\nu}N^{\mu\nu\alpha}+	\Big( -\frac{b_{2}}{2}+ 2 a_{1}+a_{2}-c_{1}\Big)N_{\mu\nu\alpha}N^{\nu\mu\alpha} \nonumber \\
		=\Big( \frac{b_{3}}{4}+4 a_{3}-c_{2} \Big)N^{(1)}_{\mu}N^{(1)}_{\nu}g^{\mu\nu}+\Big( \frac{b_{3}}{4}+a_{4}+\frac{c_{3}}{4} \Big)N^{(2)}_{\mu}N^{(2)}_{\nu}g^{\mu\nu}+a _{4}N^{(3)}_{\mu}N^{(3)}_{\nu}g^{\mu\nu} \nonumber \\
		+\Big( 2 a_{5}-\frac{c_{3}}{2}+c_{2}-\frac{b_{3}}{2}\Big) N^{(1)}_{\mu}N^{(2)}_{\nu}g^{\mu\nu}	+\Big( 2 a_{4}+\frac{c_{3}}{2}\Big)N^{(2)}_{\mu}N^{(3)}_{\nu}g^{\mu\nu}+\Big( 2 a_{5}-\frac{c_{3}}{2}\Big) N^{(1)}_{\mu}N^{(3)}_{\nu}g^{\mu\nu}
		\end{gather}
	Defining 
	\beq
	W^{\alpha}_{\;\;(\mu\nu)}=2 a_{1}Q^{\alpha}_{\;\;\mu\nu}+2 a_{2}Q_{(\mu\nu)}^{\;\;\;\;\alpha}+(2 a_{3}Q^{\alpha}+a_{5}q^{\alpha})g_{\mu\nu}+(2 a_{4}q_{(\mu} + a_{5}Q_{(\mu})\delta^{\alpha}_{\nu)}
	\eeq
	\beq
	\Pi^{\alpha\mu\nu} = c_{1}S^{\alpha\mu\nu}+c_{2}g^{\mu\nu}S^{\alpha}+c_{3}g^{\alpha\mu}S^{\nu}
	\eeq
	\beq
	A_{\mu\nu}=a_{1}(Q_{\mu\alpha\beta}Q_{\nu}^{\;\;\alpha\beta}-2 Q_{\alpha\beta\mu}Q^{\alpha\beta}_{\;\;\;\;\nu})-a_{2}Q_{\alpha\beta(\mu}Q^{\beta\alpha}_{\;\;\;\;\nu)}
	+a_{3}(Q_{\mu}Q_{\nu}-2 Q^{\alpha}Q_{\alpha\mu\nu})-a_{4}q_{\mu}q_{\nu}-a_{5}q^{\alpha}Q_{\alpha\mu\nu}
	\eeq	
	\beq
	B_{\mu\nu}=b_{1}(2S_{\nu\alpha\beta}S_{\mu}^{\;\;\;\alpha\beta}-S_{\alpha\beta\mu}S^{\alpha\beta}_{\;\;\;\;\nu})-b_{2}S_{\nu\alpha\beta}S_{\mu}^{\;\;\;\beta\alpha}+b_{3}S_{\mu}S_{\nu} 
	\eeq
	\beq
	C_{\mu\nu}=\Pi_{\mu\alpha\beta}Q_{\nu}^{\;\;\;\alpha\beta}	-( c_{1}S_{\alpha\beta\nu}Q^{\alpha\beta}_{\;\;\;\;\mu}+c_{2}S^{\alpha}Q_{\alpha\mu\nu}+c_{3}S^{\alpha}Q_{\mu\nu\alpha})=c_{1}(Q_{\mu}^{\;\;\;\alpha\beta}S_{\nu\alpha\beta}-S_{\alpha\beta\mu}Q^{\alpha\beta}_{\;\;\;\;\nu})+c_{2}(S_{\mu}Q_{\nu}-S^{\alpha}Q_{\alpha\mu\nu})
	\eeq
		 we calculate
		 \begin{gather}
		 W^{\alpha}_{\;\;\;(\mu\nu)}+\Pi^{\alpha}_{\;\;\;(\mu\nu)}=\Big( 4 a_{1}-\frac{c_{1}}{2}\Big) N_{(\mu\nu)\alpha}+\Big( 2 a_{2}+\frac{c_{1}}{2}\Big)N_{(\mu|\alpha|\nu)}+2 a_{2}N_{\alpha(\mu\nu)} \nonumber \\
		\left[ \Big( 4 a_{3}-\frac{c_{2}}{2}\Big)N_{\alpha}^{(1)}+\Big(  a_{5}+\frac{c_{2}}{2}\Big)N_{\alpha}^{(2)}+a_{5}N_{\alpha}^{(3)}\right] g_{\mu\nu} \nonumber \\
	+	g_{\alpha(\mu}  \left[ \Big( 2 a_{5}-\frac{c_{3}}{2}\Big)N_{\nu)}^{(1)}+\Big(  2 a_{4}+\frac{c_{3}}{2}\Big)N_{\nu)}^{(2)}+2 a_{4}N_{\nu)}^{(3)}\right]  
		 \end{gather}
		 as well as
		 \begin{gather}
		 A_{(\mu\nu)}+B_{(\mu\nu)}+C_{(\mu\nu)}=Q_{(\mu}^{\;\;\;\;\alpha\beta}\Big( a_{1}Q_{\nu)\alpha\beta}+c_{1} S_{\nu)\alpha\beta}\Big) -\Big( 2 a_{1}Q_{\alpha\beta(\mu}+c_{1} S_{\alpha\beta(\mu}+ a_{2} Q_{\beta\alpha(\mu}\Big) Q^{\alpha\beta}_{\;\;\;\;\nu)}\nonumber \\
		 +\Big( 2 b_{1}S_{(\mu}^{\;\;\;\alpha\beta}-b_{2} S_{(\mu}^{\;\;\;\beta\alpha}\Big) S_{\nu)\alpha\beta}-b_{1} S_{\alpha\beta\mu}S^{\alpha\beta}_{\;\;\;\nu}-Q_{\alpha\mu\nu}\Big( 2 a_{3}Q^{\alpha}+a_{5}q^{\alpha}+c_{2} S^{\alpha} \Big) \nonumber \\
		 +a_{3}Q_{\mu}Q_{\nu}-a _{4}q_{\mu}q_{\nu}+b_{3}S_{\mu}S_{\nu}+c_{2}S_{(\mu}Q_{\nu)}= \nonumber \\
		 =\Big( 2 a_{1}-\frac{c_{1}}{2}\Big) N^{\alpha\beta}_{\;\;\;\;\mu}N_{\alpha\beta\nu}+\Big( 2 a_{1}-\frac{c_{1}}{2}\Big) N^{\alpha\beta}_{\;\;\;\;\mu}N_{\beta\alpha\nu}+\frac{c_{1}}{2}N^{\alpha\beta}_{\;\;\;\;(\mu}(N_{\beta|\nu)\alpha}+N_{\alpha|\nu)\beta} ) \nonumber \\
		 -(N^{\alpha \;\;\beta}_{\;\;(\nu}+N_{(\nu}^{\;\; \alpha\beta})\left[ 2 a_{1}N_{\alpha|\mu)\beta}+ a_{2}N_{\beta|\mu)\alpha}+\Big( 2 a_{1}-\frac{c_{1}}{2}\Big)N_{\mu)\alpha\beta}+\Big( a_{2}+\frac{c_{1}}{2}\Big) N_{\mu)\beta\alpha} \right] \nonumber \\
		 +\frac{1}{4}(N_{\beta(\nu|\alpha}-N_{\beta\alpha(\nu})\Big( 2 b_{1}(N^{\beta \;\;\alpha}_{\;\;\mu)}-b_{2}N^{\alpha \;\;\beta}_{\;\;\mu)}-2 b_{1}N^{\beta\alpha}_{\;\;\;\mu)}+ b_{2}N^{\alpha\beta}_{\;\;\;\mu)}\Big)-b_{1}N_{\mu\alpha\beta}N_{\nu}^{\;\;[\alpha\beta]} \nonumber \\
		 -(N_{\mu\nu\alpha}+N_{\nu\mu\alpha})\left[ \Big( 4 a_{3}-\frac{c_{2}}{2}\Big) N^{(1)\alpha}+\Big( a_{5}+\frac{c_{2}}{2} \Big) N^{(2)\alpha}+ a_{5}N^{(3)\alpha} \right] \nonumber \\
		 +\Big( 4 a_{3}-\frac{c_{3}}{2}\Big) N_{\mu}^{(1)}N_{\nu}^{(1)}+\Big( \frac{b_{3}}{4}-a_{4}\Big)  N_{\mu}^{(2)}N_{\nu}^{(2)}-a_{4} N_{\mu}^{(3)}N_{\nu}^{(3)}\nonumber \\
		 +\Big( c_{3}-\frac{b_{3}}{2}\Big)  N_{(\mu}^{(1)}N_{\nu)}^{(2)}- 2 a_{4}  N_{(\mu}^{(2)}N_{\nu)}^{(3)}
		 \end{gather}

	We now turn our attention to the parity odd quadratic part.	In terms of the distortion, the latter	is expressed  as
		\begin{gather}
		\mathcal{L}_{odd}^{(2)}=a_{6}(N_{\alpha\mu\beta}+N_{\mu\alpha\beta})N_{\gamma \;\;\;\delta}^{\;\;\mu}\varepsilon^{\alpha\beta\gamma\delta}+(a_{6}-c_{6})N_{\alpha\mu\beta}N^{\mu}_{\;\;\gamma\delta}\varepsilon^{\alpha\beta\gamma\delta}+(a_{6}+b_{6}-c_{6})N_{\mu\alpha\beta}N^{\mu}_{\;\;\gamma\delta}\varepsilon^{\alpha\beta\gamma\delta} \nonumber \\
		+\left[ \Big( 2 c_{4}-\frac{b_{5}}{2}\Big) N_{\mu}^{(1)}+\Big( \frac{b_{5}}{2}+c_{5}\Big) N_{\mu}^{(2)}+c_{5}N_{\mu}^{(3)}\right] M^{\mu}
		\end{gather}
		Furthermore, each of its metric-variation terms (as they appear in()) can be written as
		\beq
		\varepsilon^{\alpha\beta\gamma\delta} S_{\alpha\beta\nu}S_{\gamma\delta\mu}=\varepsilon^{\alpha\beta\gamma\delta} N_{\alpha\beta\nu}N_{\gamma\delta\mu}
		\eeq
		\beq
		\varepsilon^{\alpha\beta\gamma\delta}Q_{\alpha\beta\mu}Q_{\gamma\delta\nu}=2 	\varepsilon^{\alpha\beta\gamma\delta} N_{(\alpha\mu)\beta}(N_{\gamma\nu\delta}+N_{\nu\gamma\delta})
		\eeq
		\beq
		 \varepsilon^{\alpha\beta\gamma\delta} ( 2 a_{6} Q_{\gamma\delta}^{\;\;\;\;\lambda}+c_{6} S_{\gamma\delta}^{\;\;\;\;\lambda})= (c_{6}- 2 a_{6})\varepsilon^{\alpha\beta\gamma\delta} N^{\lambda}_{\;\;\gamma\delta}- 2 a_{6} \varepsilon^{\alpha\beta\gamma\delta} N_{\gamma \;\;\;\;\delta}^{\;\;\lambda}
		\eeq
		\beq
	S_{\alpha\beta(\mu}\varepsilon^{\lambda\alpha\beta}_{\;\;\;\;\;\;\nu)}(b_{5} S_{\lambda}+c_{4}Q_{\lambda}+c_{5} q_{\lambda})=	\varepsilon^{\lambda\alpha\beta}_{\;\;\;\;\;\;(\nu}N_{\mu)\alpha\beta}\left[ \Big( 2 c_{4}-\frac{b_{5}}{2}\Big) N^{(1)}_{\lambda}+\Big(  c_{5}+\frac{b_{5}}{2}\Big) N^{(2)}_{\lambda}+ c_{5}N^{(3)}_{\lambda}\right]
		\eeq
		\beq
		S_{(\mu}t_{\nu)}=\frac{1}{2}\Big(N^{(2)}_{(\mu}-N^{(1)}_{(\mu}\Big)M_{\nu)}
		\eeq
		\beq
	t^{\alpha}Q_{(\mu\nu)\alpha} =\frac{1}{2}	\Big( N_{\nu\alpha\mu}+N_{\alpha\nu\mu}+N_{\mu\alpha\nu}+N_{\alpha\mu\nu}\Big) M^{\alpha} 
		\eeq
	With these we can also re-express ($\ref{gpart}$) solely in terms of distortion as\footnote{Also recalling that $t^{\alpha}=M^{\alpha}=N_{(4)}^{\alpha}$.}
	\begin{gather}
	\frac{1}{\sqrt{-g}}	\frac{\Big(\sqrt{-g}\mathcal{L}^{(2)}_{odd}\Big)}{\delta g^{\mu\nu}}=-b_{6} \varepsilon^{\alpha\beta\gamma\delta} N_{\alpha\beta\nu}N_{\gamma\delta\mu}+ 2 a_{6} 	\varepsilon^{\alpha\beta\gamma\delta} N_{(\alpha\mu)\beta}(N_{\gamma\nu\delta}+N_{\nu\gamma\delta}) \nonumber \\
	- g_{\beta(\mu}g_{\nu)\lambda}\frac{1}{\sqrt{-g}}(\nabla_{\alpha}- 2 S_{\alpha})\Big(\sqrt{-g}  (c_{6}- 2 a_{6})\varepsilon^{\alpha\beta\gamma\delta} N^{\lambda}_{\;\;\gamma\delta}- 2 a_{6}\sqrt{-g} \varepsilon^{\alpha\beta\gamma\delta} N_{\gamma \;\;\;\;\delta}^{\;\;\lambda} \Big)
	\nonumber \\
	-\varepsilon^{\lambda\alpha\beta}_{\;\;\;\;\;\;(\nu}N_{\mu)\alpha\beta}\left[ \Big( 2 c_{4}-\frac{b_{5}}{2}\Big) N^{(1)}_{\lambda}+\Big(  c_{5}+\frac{b_{5}}{2}\Big) N^{(2)}_{\lambda}+ c_{5}N^{(3)}_{\lambda}\right] 
	-c_{4}g_{\mu\nu}\Big(\partial_{\lambda}(\sqrt{-g}M^{\lambda})\Big)
	\nonumber \\
	-c_{5}g_{\alpha(\mu}g_{\nu)\beta}\frac{1}{\sqrt{-g}}\nabla_{\lambda}(\sqrt{-g}M^{\alpha}g^{\lambda\beta})+ c_{5} \Big(N^{(2)}_{(\mu}-N^{(1)}_{(\mu}\Big)M_{\nu)}+\frac{c_{5}}{2}	\Big( N_{\nu\alpha\mu}+N_{\alpha\nu\mu}+N_{\mu\alpha\nu}+N_{\alpha\mu\nu}\Big) M^{\alpha} 
	\end{gather}

			As for the combinations that appear on the connection field equations we readily find
		\beq
		\varepsilon^{\mu\nu\alpha\beta}Q_{\alpha\beta\lambda}=-\varepsilon^{\mu\nu\alpha\beta}N_{\alpha\lambda\beta}-\varepsilon^{\mu\nu\alpha\beta}N_{\lambda\alpha\beta}
		\eeq
		\beq
		\varepsilon^{\mu\nu\alpha\beta}S_{\alpha\beta\lambda}=\varepsilon^{\mu\nu\alpha\beta}N_{\alpha\beta\lambda}
		\eeq
		\beq
		\varepsilon_{\lambda}^{\;\;\nu\alpha\beta}Q_{\alpha\beta}^{\;\;\;\;\mu}=-\varepsilon_{\lambda}^{\;\;\nu\alpha\beta}\Big( N_{\alpha\;\;\;\;\beta}^{\;\;\mu}+N^{\mu}_{\;\;\alpha\beta} \Big)
		\eeq
		\beq
		\varepsilon_{\lambda}^{\;\;\nu\alpha\beta}S_{\alpha\beta}^{\;\;\;\;\mu}=\varepsilon_{\lambda}^{\;\;\nu\alpha\beta}N^{\mu}_{\;\;\alpha\beta}
		\eeq
			\beq
	(b_{5} S_{\lambda}+c_{4}Q_{\lambda}+c_{5} q_{\lambda})=	\left[ \Big( 2 c_{4}-\frac{b_{5}}{2}\Big) N^{(1)}_{\lambda}+\Big(  c_{5}+\frac{b_{5}}{2}\Big) N^{(2)}_{\lambda}+ c_{5}N^{(3)}_{\lambda}\right]
		\eeq
and as a result 
	\begin{gather}
	\delta_{\Gamma}\mathcal{L}_{odd}=\delta \Gamma^{\lambda}_{\;\;\;\mu\nu}\Big[ -(-2 a_{6}+c_{6})  \varepsilon^{\mu\nu\alpha\beta}N_{\alpha\lambda\beta}-(-2 a_{6}+c_{6})\varepsilon^{\mu\nu\alpha\beta}N_{\lambda\alpha\beta}                         +(2 b_{6}-c_{6})\varepsilon^{\mu\nu\alpha\beta}N_{\alpha\beta\lambda} \nonumber \\
	 +2 a_{6}\varepsilon_{\lambda}^{\;\;\nu\alpha\beta}\Big( N_{\alpha\;\;\;\;\beta}^{\;\;\mu}+N^{\mu}_{\;\;\alpha\beta} \Big)-c_{6}\varepsilon_{\lambda}^{\;\;\nu\alpha\beta}N^{\mu}_{\;\;\alpha\beta}	+\varepsilon^{\alpha\mu\nu}_{\;\;\;\;\;\;\lambda} \Big( 2 c_{4}-\frac{b_{5}}{2}\Big) N^{(1)}_{\alpha} \nonumber \\
+\Big(  c_{5}+\frac{b_{5}}{2}\Big) \varepsilon^{\alpha\mu\nu}_{\;\;\;\;\;\;\lambda} N^{(2)}_{\alpha}+ c_{5}\varepsilon^{\alpha\mu\nu}_{\;\;\;\;\;\;\lambda}N^{(3)}_{\alpha}+\Big( \frac{b_{5}}{2}+c_{5}\Big) M^{\mu}\delta_{\lambda}^{\nu}+\Big( -\frac{b_{5}}{2}+2c_{4}\Big) M^{\nu}\delta_{\lambda}^{\mu}+c_{5}g^{\mu\nu}M_{\lambda} \Big]
	\end{gather}

	\section{The $\gamma_{ij}$'s and $\alpha_{ij}$'s}
	\subsection{The $\gamma_{ij}$'s }
The relations between the elements of $\Gamma$ and the $30$ initial parameters read
\begin{gather}
\gamma_{11}=a_{1}+a_{3}+a_{71}+4 a_{81}+a_{91}\;\;, \;\; \gamma_{12}=a_{2}+a_{4}+a_{72}+4 a_{82}+ a_{92}\;\;, \;\; \gamma_{13}=a_{5}+a_{6}+a_{73}+4 a_{83}+a_{93} \nonumber \\
\gamma_{21}=a_{2}+a_{5}+4 a_{71}+ a_{81}+a_{91}\;\;, \;\; \gamma_{22}=a_{1}+a_{6}+4 a_{72}+a_{82}+ a_{92}\;\;, \;\; \gamma_{23}=a_{3}+a_{4}+4 a_{73}+ a_{83}+a_{93} \nonumber \\		
\gamma_{31}=a_{5}+a_{6}+a_{71}+ a_{81}+ 4 a_{91}\;\;, \;\; \gamma_{32}=a_{3}+a_{4}+a_{72}+ a_{82}+4 a_{92}\;\;, \;\; \gamma_{31}=a_{1}+a_{2}+a_{73}+ a_{83}+4 a_{93} \nonumber \\
\gamma_{41}=-2 (-1)^{s}\Big( b_{21}+b_{22}+b_{23}+b_{31}+b_{32}+b_{33} -3 b_{1}\Big)\; \; \; ,\; \;\; \;  \gamma_{42}=-2 (-1)^{s}\Big( b_{11}+b_{12}+b_{13}-b_{31}-b_{32}-b_{33} -3 b_{2}\Big) \nonumber \\
\gamma_{43}=2 (-1)^{s}\Big( b_{11}+b_{12}+b_{13}+b_{21}+b_{22}+b_{23}+3 b_{3} \Big)\; \;\;,\;\; \; \; \gamma_{44}=a_{1}+a_{2}+a_{3}-a_{4}-a_{5}-a_{6} \nonumber \\
\gamma_{14}=c_{1}+4 c_{2}+c_{3}-b_{12}-b_{13}-b_{21}-b_{23}-b_{31}-b_{33}\;\;, \;\; \gamma_{24}=4 c_{1}+c_{2}+ c_{3}+b_{11}+b_{22}-b_{12}-b_{21}-b_{31}-b_{32}\nonumber \\
\gamma_{34}=c_{1}+c_{2}+4 c_{3}+b_{12}-b_{13}-b_{22}+b_{23}+b_{32}-b_{33}
\end{gather}
These are the elements of the $4 \times 4$ matrix $\Gamma$.	
\subsection{The $\alpha_{ij}$'s }
The coefficients appearing in $(\ref{MasterN})$ are the first row elements of $A$, namely
\begin{gather}
\alpha_{11}=a_{1}\;\;, \;\; \alpha_{12}=a_{2}\;\;, \;\;\alpha_{13}=a_{3}\;\;, \;\;\alpha_{14}=a_{4}\;\;, \;\;\alpha_{15}=a_{5}\;\;, \;\;\alpha_{16}=a_{6}\;\;, \;\; 
\alpha_{17}=b_{11}\;\;, \;\; \alpha_{18}=b_{12}, \;\; \alpha_{19}=b_{13}\nonumber \\
\alpha_{1,10}=b_{21}\;\;, \;\; \alpha_{1,11}=b_{22}, \;\; \alpha_{1,12}=b_{23} \;\;, \;\; \alpha_{1,13}=b_{31} \;\; , \;\; \alpha_{1,14}=b_{32} \;\;,\;\; \alpha_{1,15}=b_{33}
\end{gather}
 the same goes for every other row with the last one being
\begin{gather}
\alpha_{15,1}=(-1)^{s}(2 b_{11}-b_{12}-b_{13}) \;\;, \;\; \alpha_{15,2}=(-1)^{s}(2 b_{31}-b_{32}-b_{33}) \;\;, \;\; \alpha_{15,3}=-(-1)^{s}(2 b_{21}-b_{22}-b_{23}) \nonumber \\
\alpha_{15,4}=(-1)^{s}(2 b_{21}-b_{22}-b_{23}) \;\;, \;\; \alpha_{15,5}=-(-1)^{s}(2 b_{31}-b_{32}-b_{33}) \;\;, \;\; \alpha_{15,6}=-(-1)^{s}(2 b_{11}-b_{12}-b_{13}) \nonumber \\
\alpha_{15,7}=0\;\;, \;\; \alpha_{15,8}=a_{3}-a_{5} \;\;, \;\; \alpha_{15,9}=0 \;\;, \;\; \alpha_{15,10}=0 \;\;, \;\; \alpha_{15,11}=a_{6}-a_{2} \;\;, \;\; \alpha_{15,12}=0 \nonumber \\
\alpha_{15,3}=0\;\;, \;\; \alpha_{15,14}=a_{1}-a_{4} \;\;, \;\; \alpha_{15,15}=0 
\end{gather}

	\bibliographystyle{unsrt}
	\bibliography{ref}

\end{document}